\documentclass[fleqn]{mn2e}
\usepackage{amssymb}
\usepackage{amsmath}
\usepackage{graphicx}
\usepackage{psfrag}

\newcommand{\bfe}{{\boldsymbol{e}}}
\newcommand{\bfk}{{\boldsymbol{k}}}

\newcommand{\bfr}{{\boldsymbol{r}}}
\newcommand{\bfv}{{\boldsymbol{v}}}
\newcommand{\txd}{{\text{d}}}
\newcommand{\nodu}{{\tt nodu}}
\newcommand{\abso}{{\tt abso}}
\newcommand{\dust}{{\tt dust}}
\renewcommand{\leq}{\leqslant}
\renewcommand{\geq}{\geqslant}

\title[Kinematics of dusty ellipticals III]{Kinematics of elliptical
galaxies with a diffuse dust component -- III.\ A Monte Carlo
approach to include the effects of scattering}

\author[Baes \& Dejonghe]{Maarten Baes\thanks{Postdoctoral Fellow of
the Fund for Scientific Research, Flanders, Belgium
(F.W.O.-Vlaanderen)} and Herwig Dejonghe \\Sterrenkundig
Observatorium, Universiteit Gent, Krijgslaan 281-S9, B-9000 Gent,
Belgium, maarten.baes@rug.ac.be}

\begin{document}

\maketitle

\begin{abstract}
This paper is the third one in a series, intended to investigate
how the observed kinematics of elliptical galaxies are affected by
dust attenuation. In Paper~I and Paper~II, we investigated the
effects of dust absorption; here we extend our modelling in order
to include the effects of scattering. We describe how kinematical
information can be combined with the radiative transfer equation,
and present a Monte Carlo code that can handle kinematical
information in an elegant way.

Compared to the case where only absorption is taken into account,
we find that dust attenuation considerably affects the observed
kinematics when scattering is included. For the central lines of
sight, dust can either decrease or increase the central observed
velocity dispersion. The most important effect of dust
attenuation, however, is found at large projected radii. The
kinematics at these lines of sight are strongly affected by
photons scattered into these lines of sight, which were emitted by
high-velocity stars in the central regions of the galaxy. These
photons bias the LOSVDs towards high line-of-sight velocities, and
significantly increase the observed velocity dispersion and LOSVD
shape parameters. These effects are similar to the expected
kinematical signature of a dark matter halo, such that dust
attenuation may form an alternative explanation for the usual
stellar kinematical evidence for dark matter halos around
elliptical galaxies.

We apply our results to discuss several other topics in galactic
dynamics, where we feel dust attenuation should be taken into
account. In particular, we argue that the kinematics observed at
various wavelengths can help to constrain the spatial distribution
of dust in elliptical galaxies.
\end{abstract}

\begin{keywords}
dust, extinction -- galaxies: elliptical and lenticular, cD --
galaxies: kinematics and dynamics -- radiative transfer --
scattering
\end{keywords}

\section{Introduction}

It is now generally accepted that elliptical galaxies are
complicated objects, containing a variety of stellar populations,
with a still poorly constrained dark matter content and a
multi-component interstellar medium. The latter, and in particular
the interstellar dust component, is not well understood.
Nevertheless, knowledge about the dust component in galaxies is of
fundamental importance. On the one hand, interstellar dust plays
an active role in e.g.\ interstellar chemistry and star formation,
and is therefore a key ingredient to understand galaxy structure
and evolution. On the other hand, dust is very effective in
absorbing and scattering UV and optical light, and knowledge about
its presence, quantity and properties is necessary to correctly
interpret any observable.

The far-infrared fluxes detected in the eighties by the
{\em{IRAS}} satellite at 60 $\mu$m and 100 $\mu$m unveiled the
presence of a substantial amount of dust in elliptical galaxies
(Jura 1986; Bally \& Thronson 1989; Knapp, Gunn \& Wynn-Williams
1989). At that time, dust was already observed in the form of dust
lanes and patches for a number of early-type galaxies (Hawarden et
al.\ 1981; Ebneter \& Balick 1985; V\'eron-Cetty \& V\'eron 1988).
Recent surveys indicate that dust extinction features are present
in a large fraction of early-type galaxies (van Dokkum \& Franx
1995; Ferrari et al. 1999; Tomita et al. 2000; Tran et al. 2001).
These dust features, however, can not account for the high
{\em{IRAS}} fluxes: the dust masses estimated from the {\em{IRAS}}
measurements exceed those estimated from integrating the
extinction features by nearly an order of magnitude (Goudfrooij \&
de Jong 1995). Moreover, the {\em{IRAS}} dust mass estimates are a
lower limit for the true dust masses, because {\em{IRAS}} is not
sensitive to cold dust, which emits the bulk of its radiation
longwards of 100 $\mu$m. Calculations with a more realistic dust
temperature distribution (Merluzzi 1998) and submillimeter
observations (Fich \& Hodge 1993; Wiklind \& Henkel 1995) may
indicate dust masses up to a magnitude higher than the {\em{IRAS}}
estimates. The only plausible way to solve this dust mass
discrepancy is to assume that the interstellar dust in elliptical
galaxies exists as a two-component medium: the less massive
component is optically visible in the form of dust lanes, whereas
the more massive one is distributed diffusely over the galaxy
(Goudfrooij \& de Jong 1995).

Such a diffusely distributed dust distribution can be traced in
two ways. The most obvious way is to map the thermal emission of
the dust with spatially resolved far-infrared or submillimeter
observations. A first effort to do this was undertaken by Leeuw,
Sansom \& Robson (2000), who used {\em{SCUBA}} imaging at 850
$\mu$m to look for thermal emission from dust in NGC\,4374.
Unfortunately, they could not detect a diffuse dust component, and
found that the bulk of the 850 $\mu$m core emission is more likely
the result of synchrotron radiation. Deeper imaging covering a
larger wavelength range is necessary to constrain the spatial
distribution of dust this way. Hopefully, the new generation of
far-infrared and submillimeter instrumentation such as
{\em{SIRTF}} and {\em{ALMA}} will help to clarify this issue.

A second way in which the diffuse dust component in elliptical
galaxies can be traced is by colour gradients: because the
extinction efficiency of dust grains decreases with wavelength, a
diffuse dust component is expected to generate bluer colours at
larger projected radii. Goudfrooij \& de Jong (1995) and Wise \&
Silva (1996) found that the dust distributions, necessary to
create the colour gradients observed in a sample of elliptical
galaxies, are in reasonable agreement with their observed
integrated FIR fluxes. However, various arguments seriously
complicate the interpretation of colour gradients. Foremost, dust
attenuation\footnote{We refer to dust attenuation as the combined
effect of absorption and scattering.} is not the only process that
generates colour gradients: also age and metallicity variations
can cause broadband colour gradients. Recently, Michard (2000)
argued that the mean observed colour gradient ratios of a sample
of elliptical galaxies are more likely the result of metallicity
than of diffuse dust. And even if the dust attenuation were the
only process responsible for the generation of colour gradients,
tracing and quantifying the amount of dust would still be
complicated. First, the dust is not located between the source and
the observer, but well mixed with the stars. As a result, the
amount of dust is not simply proportional to the amount of
reddening (Disney, Davies \& Phillipps 1989). Second, bluing due
to scattering partly compensates the effects of reddening due to
absorption, which suppresses the formation of large broad band
colour gradients even if a substantial amount of dust is present
(Witt, Thronson \& Capuano 1992).

We have set up a program to investigate the effects of diffuse
dust on the observed kinematics of elliptical galaxies. In the
first two papers of this series (Baes \& Dejonghe 2000, hereafter
Paper~I; Baes, Dejonghe \& De Rijcke 2000, hereafter Paper~II) we
investigated how the light profile and the observed kinematics of
elliptical galaxies are affected by dust absorption. In this
series we extend our models by including the effect of scattering.
In Section \ref{RTE.sec} we explain how the processes of dust
absorption and scattering can affect the observables of galaxies.
In particular, we explain how kinematical information can be
included in the radiative transfer equation. In Section
\ref{montecarlo.sec}, we argue that a Monte Carlo method is the
most straightforward way to do this. We describe the code we
developed, with a special emphasis on the calculation of the
observed kinematics. We use this code to investigate the effects
of dust attenuation on the observed kinematics in elliptical
galaxies. In Section \ref{galaxymodels.sec} we present a set of
simple elliptical galaxy models, consisting of a stellar and a
dust component. We demonstrate how dust attenuation affects the
light profile and observed kinematics of these models in Section
\ref{results.sec}. Section \ref{discussion.sec} is devoted to a
discussion of our results, and finally, Section
\ref{conclusions.sec} sums up.

\section{Radiative transfer in dusty galaxies}
\label{RTE.sec}

\subsection{The general radiative transfer equation}
\label{genRTE.sec}

The basis for any study of attenuation is the radiative transfer
equation (RTE), which statistically describes the interaction
between matter and light. In a general form, the time-independent
RTE can be written as (Chandrasekhar 1960; Mihalas 1978)
\begin{equation}
    \frac{\txd I}{\txd s}(\bfr,\bfk)
    =
    j(\bfr,\bfk)
    -
    \kappa(\bfr)\,I(\bfr,\bfk),
\label{RTE_gen}
\end{equation}
where $s$ is the path length, and $I(\bfr,\bfk)$ represents the
intensity at a position $\bfr$ into a direction $\bfk$. The
right-hand side of the equation contains two terms that represent
how the radiation field changes as a result of interactions with
the matter. The first term, the total emissivity $j(\bfr,\bfk)$
accounts for the sources of the radiation field. The second term,
where $\kappa(\bfr)$ is the total opacity coefficient, represents
the sinks. The specific form of both terms depends on which
physical processes are taken into account, (e.g.\ stellar
emission, absorption, scattering) and they can be complicated
functions that depend on the intensity itself.

In a galaxy without dust attenuation, there are no sink terms, and
the only source in the radiation field is the (isotropic) emission
of photons by stars, such that $j(\bfr,\bfk)=\ell(\bfr)$ where
$\ell(\bfr)$ represents the stellar emissivity. If we take
absorption by dust grains into account, a sink term must be added,
which accounts for the loss of photons from the radiation
field.\footnote{In fact, dust absorption will also account for an
additional source term, because the energy absorbed by grains will
be re-emitted at infrared wavelengths. Because we are primarily
interested in the optical and near-infrared regimes, however, this
source term can safely be neglected.} The total opacity then
equals the absorption coefficient, i.e.\
$\kappa(\bfr)=\kappa_{\text{abs}}(\bfr)$. In either of these
situations, the RTE is a ordinary differential equation, and it
can be solved by an integration along the line of sight.

Dust grains do not only absorb photons, they also scatter them,
i.e.\ a number of photons are, as a result of an interaction with
a dust grain, removed from their path and sent into another
direction. This physical process is not a rare phenomenon: for
typical Milky Way dust grains, the probability for scattering even
slightly exceeds the probability for absorption in the optical
wavelength range (Table \ref{dustprop.tab}). It is therefore
obvious that scattering should be included in radiative transfer
calculations. This will add two extra terms to the RTE. The first
one is a sink term that accounts for the loss of photons scattered
out of the beam. The total opacity will hence be the sum of the
absorption coefficient and the scattering coefficient,
$\kappa(\bfr) = \kappa_{\text{abs}}(\bfr) +
\kappa_{\text{sca}}(\bfr)$. The second extra term is a source term
that characterizes the gain of photons scattered into the beam.
More precisely, this term will contain the contribution of photons
that had another direction $\bfk'$, but are now scattered into the
direction $\bfk$. Generally, the distribution of angles after a
scattering process are described by a scattering phase function
$\Phi(\bfr,\bfk,\bfk')$, which describes the probability that a
photon which comes from the direction $\bfk'$ and is scattered at
$\bfr$, will have $\bfk$ as its new direction. By convention, it
is normalized as
\begin{equation}
    \iint\frac{\txd\Omega'}{4\pi}\,\Phi(\bfr,\bfk,\bfk')=1
    \qquad\text{for all $\bfr$ and all $\bfk$.}
\end{equation}
The fraction of the intensity scattered from a solid angle
$\txd\Omega'$ around an arbitrary direction $\bfk'$ into the
direction $\bfk$ will therefore equal $\kappa_{\text{sca}}(\bfr)\,
I(\bfr,\bfk')\, \Phi(\bfr,\bfk,\bfk')\, \txd\Omega' / 4\pi$, such
that the extra source term that has to be added to the RTE reads
\begin{equation}
    j_{\text{sca}}(\bfr,\bfk)
    =
    \kappa_{\text{sca}}(\bfr)
    \iint\frac{\txd\Omega'}{4\pi}\,
    I(\bfr,\bfk')\, \Phi(\bfr,\bfk,\bfk').
\end{equation}
It is convenient to introduce the scattering albedo $\omega$ as
the ratio of the scattering coefficient to the total opacity
coefficient,
\begin{equation}
    \omega(\bfr)
    =
    \frac{\kappa_{\text{sca}}(\bfr)}{\kappa(\bfr)}
    =
    1-\frac{\kappa_{\text{abs}}(\bfr)}{\kappa(\bfr)}.
\end{equation}
We finally find for the RTE,
\begin{multline}
    \frac{\txd I}{\txd s}(\bfr,\bfk)
    =
    \ell(\bfr) - \kappa(\bfr)\,I(\bfr,\bfk)
    \\
    +
    \omega(\bfr)\,\kappa(\bfr)\,\iint\frac{\txd\Omega'}{4\pi}\,
    I(\bfr,\bfk')\,\Phi(\bfr,\bfk,\bfk').
\label{RTE_dust}
\end{multline}
The inclusion of scattering turns the RTE into a
integro-differential equation, far more complicated than the
ordinary differential equation if only absorption is taken into
account. In particular, the scattering term is responsible for the
coupling of the RTE along different paths: due to the integration
over the angle, we cannot solve the RTE for a single path, but we
have to solve it for all paths at the same time.

\subsection{The geometry}

The complexity of the RTE does not only depend on which physical
processes are taken into account (the right-hand side), but also
on the geometry of the system. Indeed, the path length appearing
in the left-hand side of the RTE is in general a function of
position and direction, i.e.\ $s = s(\bfr,\bfk)$. As a
consequence, the (time-independent) RTE is a partial differential
equation with five independent coordinates (see e.g.\ Mihalas
1978). This complexity, however, is reduced if the system has
symmetries. For example, in an axially symmetric geometry, the
azimuthal dependence vanishes, such that only four independent
coordinates need to be considered.

Particularly interesting is the spherical geometry, because in
that case only two coordinates remain: the radius $r$ and the
cosine $\mu$ of the angle between the direction $\bfk$ and the
local radial direction. The aim of this paper is to provide a
global picture of the effects of dust attenuation on the observed
kinematics of elliptical galaxies, rather than modelling a
specific, possibly geometrically complex object. For this goal,
the assumption of spherical symmetry is satisfactory.

There are no ways to solve the general RTE (\ref{RTE_dust})
analytically, but several techniques have been developed to solve
it numerically. And because many astrophysical systems can in a
first approximation be considered spherically symmetric, many
efforts have yet been spent on the RTE in a spherical geometry.
The pioneering work started nearly 70 years ago (Kosirev 1934;
Chandrasekhar 1934), and, in particular since the 1970s, a vast
number of different approaches has been developed (e.g.\ Hummer \&
Rybicki 1971; Schmid-Burgk 1975; Witt 1977; Flannery, Roberge \&
Rybicki 1980; Yorke 1980; Rowan-Robinson 1980; Rogers \& Martin
1984; Peraiah \& Varghese 1985; Gros, Crivellari \& Simonneau
1997). Most of these techniques, however, are not suitable for our
needs, because it seems hard (and at least not obvious) to extend
them such that they can handle kinematical information. Only for
Monte Carlo methods, the kinematical information could be included
in an elegant way.

\subsection{Including kinematical information}

As long as scattering is not taken into account, the inclusion of
velocity information in the RTE is rather straightforward. Indeed,
instead of taking into account the entire stellar emissivity, we
can just consider the light from those stars whose velocity
component in the direction of the observer equals $v_\parallel$.
The emissivity of these stars is given by
$\ell(\bfr)\,\phi(\bfr,\bfk,v_\parallel)$, where
$\phi(\bfr,\bfk,v_\parallel)$ is the spatial LOSVD, i.e.\ the
probability for a star at position $\bfr$ to have a velocity
component $v_\parallel$ in the direction $\bfk$ (Appendix~A).
Solving the RTE with this emissivity, we obtain the LOSVDs
observed in the plane of the sky. Hence, if either dust
attenuation is completely neglected or only absorption by dust
grains is taken into account, the observed LOSVDs can be
calculated from the spatial LOSVDs through a single integration
along the line of sight. Similar relations hold between the
moments of the distribution function, and the moments of the
LOSVDs, e.g.\ the observed velocity dispersion profile can be
determined from the intrinsic velocity dispersions by a simple
integration along the line of sight. For more details, we refer to
Paper~I.

\begin{figure}
\psfrag{*}{\small star}
\psfrag{k0}{\small $\bfk_0$}
\psfrag{d}{\small dust grain}
\psfrag{kobs}{$\bfk_{\text{obs}}$}
\psfrag{vsys}{$\bfv_{\text{sys}}$}
\psfrag{vd}{$\bfv_d$}
\psfrag{vs}{$\bfv_*$}
\psfrag{observer}{observer}
\centering
\includegraphics[clip,width=7cm]{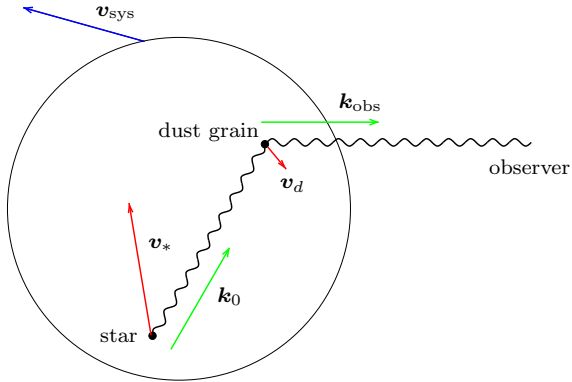}
\caption{The inclusion of kinematical information in the radiative
transfer equation. This figure shows the trajectory of a photon
through the galaxy (undulating line): after being emitted by the
star, it is scattered once before it leaves the galaxy. See text
for more details.}
\label{kininfo.eps}
\end{figure}

When scattering is included, however, the inclusion of kinematical
information becomes much more complicated, because the velocities
of both the stars and the dust grains that scatter their light
need to be taken into account. Consider, for example, Fig.\
\ref{kininfo.eps}. A photon with rest wavelength $\lambda_0$ is
emitted by a star moving with a velocity $\bfv_*$ into a direction
$\bfk_0$ (in a reference frame centered on the galaxy center).
Assume this photon is scattered by a dust grain moving with
velocity $\bfv_d$, and then scattered towards the observer. The
wavelength "detected" by the dust grain is determined by the
relative velocity between star and dust grain, i.e.\ the dust
grain detects a wavelength
\begin{equation}
    \lambda
    =
    \lambda_0 \left[1-\frac{(\bfv_*-\bfv_d)\cdot\bfk_0}{c}\right].
\end{equation}
The dust grains scatters the photon coherently into the direction
$\bfk_{\text{obs}}$, such that the wavelength detected by the
observer equals
\begin{equation}
    \lambda
    =
    \lambda_0 \left[1-\frac{(\bfv_*-\bfv_d)\cdot\bfk_0}{c}\right]
    \left[1-\frac{(\bfv_d+\bfv_{\text{sys}})\cdot\bfk_{\text{obs}}}{c}\right],
\label{lam1}
\end{equation}
where $\bfv_{\text{sys}}$ is the system velocity with respect to
the observer. The line-of-sight velocity $v_\parallel$ detected by
the observer can be found from the relation
\begin{equation}
    \lambda
    =
    \lambda_0\,
    \left[1-\frac{v_\parallel+
    \bfv_{\text{sys}}\cdot\bfk_{\text{obs}}}{c}\right],
\label{lam2}
\end{equation}
which defines $v_\parallel$ and already incorporates the system
velocity. Because the velocities of dust grains and stars in
galaxies are very small compared the speed of light, second order
terms in $v/c$ can be safely neglected, such that with
(\ref{lam1}) and (\ref{lam2}),
\begin{equation}
    v_\parallel
    =
    \bfv_*\cdot\bfk_0 +
    \bfv_d\cdot\left(\bfk_{\text{obs}}-\bfk_0\right).
\end{equation}
When more scattering events are involved in the photon's
trajectory, the relative velocities of each pair of subsequent
dust grains have to be taken into account. If we denote the total
number of scattering events with $M$, the velocities of the dust
grains with $\bfv_{d_i}$ and the propagation directions with
$\bfk_i$ (for $i=1\ldots M$), we obtain
\begin{equation}
    v_\parallel
    =
    \bfv_*\cdot\bfk_0
    +
    \sum_{i=1}^{M} \bfv_{d_i}\cdot\left(\bfk_i-\bfk_{i-1}\right).
\label{VVV}
\end{equation}
In general, therefore, the inclusion of kinematical information
into radiative transfer problems is very complex. For our problem
of dust attenuation in elliptical galaxies, however, we can make
one assumption that reduces the complexity considerably: the
summation in formula (\ref{VVV}), i.e.\ the common contribution of
the dust velocities, is in general negligible with respect to the
first term, the contribution of the stellar velocity. Several
arguments support this assumption. [1]~It is reasonable to assume
that the dust grains have smaller velocities than the stars in
elliptical galaxies. Indeed, if the dust grains were to have large
velocities, they would collide and heat up. However, the lion's
share of the dust in elliptical galaxies is assumed to be cold
(Fich \& Hodge 1993; Goudfrooij \& de Jong 1995; Wiklind \& Henkel
1995; Merluzzi 1998). [2]~Scattering off interstellar dust is
generally anisotropic, with a larger probability for forward
scattering (see Section \ref{phafu.sec}). In Appendix~B we show
that the anisotropic nature of scattering contributes to reducing
the importance of the dust grain velocity terms in equation
(\ref{VVV}). [3]~There is no reason why the velocity of dust
grains in an elliptical galaxy would have a preferential
direction. The individual terms in the summation in formula
(\ref{VVV}) will therefore have a random sign, and the summation
will in the mean be washed out by multiple scattering.

As a consequence, we can write equation (\ref{VVV}) simply as
$v_\parallel = \bfv_*\cdot\bfk_0$. Each photon hence carries the
velocity component of the star in the direction of the emission of
the photon.

\section{Description of the Monte Carlo routine}
\label{montecarlo.sec}

\subsection{Basic characteristics}

We constructed a Monte Carlo code to solve our radiative transfer
problem. Usually, the main argument against Monte Carlo methods is
that it is computationally rather expensive compared to other
methods (see e.g.\ Baes \& Dejonghe 2001a). It has other
advantages, however, which make it very competitive, in particular
in an era when CPU time is not the most stringent limitation
anymore. Important qualities of the Monte Carlo method are the
possibility for a proper error analysis and a very wide
flexibility. Nice examples of this flexibility are the ability to
handle arbitrary geometries (Wolf, Fischer \& Pfau 1998; Wood \&
Reynolds 1999; Gordon et al. 2001), the polarization of the
scattered radiation (Code \& Whitney 1995; Bianchi, Ferrara \&
Giovanardi 1996), the clumpiness of the interstellar medium (Witt
\& Gordon 1996; Bianchi et al.\ 2000), and the self-consistent
heating and re-emission of the absorbed radiation (Wolf, Henning
\& Stecklum 1999; Bianchi, Davies \& Alton 2000; Misselt et al.\
2001). We make use of another possibility of the Monte Carlo
method: its ability to include kinematical information in an
elegant way. This possibility has recently been explored by
Matthews \& Wood (2001), who studied the effect of dust
attenuation on the observed rotation curve in the spiral galaxies.
Our modelling opens up the possibility to construct all LOSVDs and
hence to investigate the entire observed kinematical structure of
galaxies.

The principles of the Monte Carlo technique are outlined in detail
by various authors (Cashwell \& Everett 1959; Mattila 1970; Witt
1977; Yusef-Zadeh, Morris \& White 1984; Fischer, Henning \& Yorke
1994; Bianchi et al.\ 1996). Basically, the routine consists of
following the individual trajectory of a very large number of
photons through the galaxy. A photon's history is given by a
number of quantities such as the position and propagation
direction at birth, the distance the photon travels before it
interacts with a dust grain, the kind of this interaction, etc.
Each of these quantities is described statistically by a random
variable, taken from a particular probability density. Our
approach is based on Witt (1977) and Bianchi et al.\ (1996). It
includes the use of a continuous three-dimensional cartesian
reference system (no grid). Furthermore, we use the classical
tricks to optimize the routine: we assign a weight to each photon
in order to avoid the loss of photons due to absorption, and we
apply the forced first scattering mechanism to improve the
statistics of the scattered radiation (Cashwell \& Everett 1959;
Witt 1977). The novelty of our Monte Carlo code is that it can
calculate both photometric and kinematic data: not only the light
profile, but also the LOSVDs and the projected kinematic moments.
It simultaneously calculates these data in three modes: without
dust attenuation, with only absorption taken into account and with
dust attenuation fully taken into account. It is a monochromatic
code, i.e.\ it calculates the observables in one single
wavelength. If desired, the wavelength dependence of the results
can be investigated by running the code various times at different
wavelengths. We implemented the central wavelengths of the optical
$U$, $B$, $V$, $R$, $I$ and the near-infrared $J$, $H$ and $K$
broadbands.

\subsection{Calculation of the light profile}

The light profile of the galaxy is constructed by simulating an
imaging process. If a photon leaves the galaxy, we record the
position of the last scattering event and the final propagation
direction of the photon. Because in spherical symmetry, a path can
be determined by its projected radius, we only need to record the
projected radius of the photon's final path. The registration of
the photons is done by constructing a histogram of the photons
leaving the galaxy as a function of $R$, where the contribution of
each photon is measured by its weight (see Witt 1977).

\subsection{Calculation of the LOSVDs}

The Monte Carlo method provides an elegant possibility to include
kinematical information. Indeed, we argued that the velocity
information that a photon carries is essentially the velocity
component of the star in the direction of the emission. This extra
information can easily be included in the Monte Carlo routine. The
only extension is that we do not only have to generate an initial
position and propagation direction for each photon, but also a
stellar velocity. The initial position and the stellar velocity
have to be extracted in accordance with the phase space
distribution function $F(\bfr,\bfv)$, which represents the
probability of finding a star at the position $\bfr$ with velocity
$\bfv$. A practical way to do so, is following the strategy of
Wybo \& Dejonghe (1996), who generated sets of phase space
coordinates of stars to create $N$-body representations of
globular clusters. First, a random position $\bfr_0$ is determined
from the stellar emissivity, and then a random stellar velocity
$\bfv_*$ is generated from the three-dimensional probability
density $p(\bfv) = {\mathcal{L}}\,F(\bfr_0,\bfv) / \ell(\bfr_0)$,
using the acceptance-rejection method (Abramowitz \& Stegun 1965;
Press et al.\ 1989). Next, we determine a random emission
direction $\bfk_0$, and determine the velocity component of
$\bfv_*$ in the direction $\bfk_0$, i.e.\
$v_\parallel=\bfv_*\cdot\bfk_0$. Given these initial conditions,
we follow the photon until the exit conditions are satisfied, and
calculate the projected radius of the final line of sight. The
photon will then be stacked, according to its weight, in the
appropriate bin in a two-dimensional array with axes $R$ and
$v_\parallel$. This process is repeated this for a large number of
photons and two-dimensional histograms are formed.

This method can be optimized in two ways. Firstly, we see that the
velocity vector $\bfv_*$ does not show up in the Monte Carlo
procedure itself, but only in the classification process.
Moreover, we actually don't need the entire information contained
in the velocity $\bfv_*$: we only need the component $v_\parallel$
of the velocity in the direction $\bfk_0$. Therefore it is
sufficient to generate a random $v_\parallel$ from the spatial
LOSVD $\phi(\bfr_0,\bfk_0,v_\parallel)$, which represents the
probability density of line-of-sight velocities for a star at
position $\bfr_0$ in the direction $\bfk_0$. Instead of generating
a three-dimensional velocity vector for each photon in the
beginning of the Monte Carlo cycle, it is hence sufficient to
generate one single line-of-sight velocity component. This saves
many random number generations and distribution function
evaluations, which are costly processes.

A second optimization eliminates the generation a $v_\parallel$
altogether. Instead of generating one single random $v_\parallel$,
we can, for each bin in the velocity direction, calculate the
probability that the line-of-sight velocity will fall in that bin.
If the boundaries of a bin are given by $v_{\parallel,j-1}$ and
$v_{\parallel,j}$, this probability equals
\begin{multline}
    \int_{v_{\parallel,j-1}}^{v_{{\parallel,j}}}
    \phi(\bfr_1,\bfk_0,v_\parallel)\,\txd v_\parallel
    \\
    \approx
    \frac{1}{2}\,
    \frac{\phi(\bfr_1,\bfk_0,v_{\parallel,j-1})
    +\phi(\bfr_1,\bfk_0,v_{\parallel,j})}
    {v_{\parallel,j}-v_{\parallel,j-1}}.
\end{multline}
For each photon, we assign a weighted value to each velocity bin
corresponding to the final projected radius, which gives better
statistics than dropping the photon in a single bin.

\subsection{Calculation of the projected kinematics}

Of course, we are not able anymore to write down a direct
connection between the intrinsic moments of the distribution
function and the projected moments on the plane of the sky,
contrary to the absorption-only case (Paper~I). Instead, we will
calculate the projected moments directly from the obtained LOSVDs.
The projected kinematics we consider are the mean projected
velocity $\bar{v}_p$, the projected velocity dispersion
$\sigma_p$, the skewness $k_3$, the kurtosis $k_4$, and the lowest
order Gauss-Hermite coefficients $h_3$ and $h_4$. Note, however,
that for spherically symmetric non-rotating galaxies all odd
moments vanish, i.e.\ $\bar{v}_p=k_3=h_3=0$.

\subsection{Error analysis}

Obviously, a reliable error analysis is necessary to estimate the
accuracy of the results. But it will be even more important in a
later stage. Indeed, we are presently incorporating this radiative
transfer mechanism into the QP program (Dejonghe 1989), a code for
the dynamical modelling of gravitating systems. Therefore, it is
necessary that we can dynamically evaluate the error bars on any
observable: the program has to be able to calculate any observable
with an preset accuracy, required by either the user or another
part of the code itself. Monte Carlo methods satisfy this need:
they can keep on adding photons until any accuracy requirement is
satisfied.

The determination of the errors is easy for observables that are
directly proportional to the number of photons detected in the
bins, e.g.\ the intensity. Because of the Poisson character of the
noise, the error bars on these quantities can be directly
calculated from the square root of the number of photons. The
observed kinematics we calculate, however, are not directly
proportional to number of photons obtained in the bins, such that
we cannot adopt this simple procedure. Therefore, we estimate the
error on our observables using an alternative method. Assume we
have calculated a value $f$ for a given observable (either
photometrical or kinematical) after a run with $N$ photons. We
divide the total number of photons in $M$ subsets, and for each of
these subsets, we calculate the corresponding observable $f_i$.
The uncertainty on $f$ is estimated as,
\begin{equation}
    \Delta f = \left[\frac{1}{M}\sum_{i=1}^M
    (f_i-f)^2\right]^{1/2}.
\end{equation}
Typically, we set $M=100$, and we adopt a minimum number of $10^6$
photons, i.e.\ $10^4$ photons per subset.

\section{The galaxy models}
\label{galaxymodels.sec}

To investigate the effects of dust attenuation on the light profile
and the observed kinematics, we will adopt a set of dusty galaxy
models, similar to those in Paper~I and Paper~II. They are spherically
symmetric, and consist of a stellar and a dust component.

\subsection{The stellar component}

For the stellar component we use a self-consistent Plummer model
(Plummer 1911; Dejonghe 1987), defined by the potential-density
pair
\begin{subequations}
\begin{gather}
    \psi(r)
    =
    \frac{GM}{\sqrt{r^2+c^2}},
    \\
    \rho(r)
    =
    \frac{3M}{4\pi}\,\left(1+\frac{r^2}{c^2}\right)^{-5/2}.
\end{gather}
\end{subequations}
We adopt a core radius $c=5$\,kpc, a total mass
$M=6\times10^{11}$\,M$_\odot$ and a (constant) mass-to-light ratio
$\Upsilon=4$\,$\Upsilon_\odot$, such that the total emitted
luminosity equals ${\mathcal{L}} =
1.5\times10^{11}$\,L$_\odot$.\footnote{These parameters differ
from those adopted in Paper~II, which contains an error: the
adopted mass is inconsistent with the velocity dispersion. When we
refer to results of Paper~II, these are scaled to the parameters
adopted in this paper.} As in Paper~I, we consider different
Plummer models, each with a different value for the parameter $q$,
which describes the internal anisotropy of the stellar orbits. We
adopt three different models: a radial model ($q=2$), the
isotropic model ($q=0$) and a tangential model ($q=-6$). These are
the same values as in Paper~I, except that we replace the radial
model with $q=1$ by the model with $q=2$. The reason for
preferring $q=2$ above $q=1$ has a computationally nature: the
spatial LOSVDs can be calculated analytically for $q=2$, whereas
this is not possible for the $q=1$ model (Appendix~A).

\subsection{The dust component}

As shown in Section \ref{genRTE.sec}, a complete characterization
of the dust component requires a specification of the opacity
$\kappa(r)$, the albedo $\omega(r)$ and the scattering phase
function $\Phi(r,\bfk,\bfk')$. The opacity sets the total amount
of dust and its spatial distribution, whereas the other two
quantities describe the optical properties at a given distance to
the galaxy centre. In order to limit the number of parameters in
our model, we will assume that the optical properties of the dust
grains are the same all over the galaxy, such that the albedo and
the phase function become independent of $r$. We will return to
this assumption in Section 5.3.3.

\subsubsection{The dust distribution}

We adopt the same family of dust components considered in Paper~I
and Paper~II, characterized by the dust number density
\begin{equation}
    n(r)
    =
    n_0\left(1+\frac{r^2}{c^2}\right)^{-\alpha/2},
\end{equation}
where $n_0$ is the dust density at the center of the galaxy, and
$c$ and $\alpha$ determine the actual shape of the dust
distribution. Because the effect of varying $\alpha$ and $c$ are
roughly similar, we restrict ourselves to varying the dust
exponent $\alpha$, and we fix the dust core radius to $c=5$~kpc,
equal to the stellar core radius. The variation of $\alpha$ still
allows us to explore a large variety of star-dust geometries:
small values of $\alpha$ correspond to centrally concentrated
distributions, whereas for large $\alpha$ the distribution is
shallower (Paper~I).

At a fixed wavelength $\lambda$, the opacity function
$\kappa_\lambda(r)$ can be derived from the dust density by
multiplication with the (wavelength-dependent) dust cross-section
$s_\lambda$. To characterize the dust distribution at a given
wavelength, we thus need three parameters: the dust exponent
$\alpha$, the central dust number density, and the dust
cross-section $s_\lambda$. It is more convenient, however, to use
another, equivalent, set of parameters. Indeed, if the opacity
function is known at one particular wavelength, it can be
determined at any other wavelength. Usually, the {\em V}-band is
used as the reference wavelength. We can then write
$\kappa_\lambda(r) = X_\lambda\,\kappa_V(r)$, where $X_\lambda
\equiv s_\lambda/s_V$ is the extinction coefficient, relative to
the {\em V} band. It is then customary to introduce the total {\em
V}-band optical depth as\footnote{In this definition, $\tau_V$ is
defined as the optical depth from the center of the galaxy to the
edge. We want to stress that this definition differs a factor 2
with the definition of the optical depth we adopted in Paper~I and
Paper~II, where we defined $\tau_V$ as the total optical depth
along the central line of sight, i.e.\ from one edge to the other.
The definition of $\tau_V$ we adopt here, however, is more
commonly used, and it will allow the reader to link the obtained
results more easily to those found in the literature (e.g.\ Witt
et al.\ 1992; Wise \& Silva 1996). When we refer to the results of
Paper~I and Paper~II, we will adopt the convention used here.}
\begin{equation}
    \tau_V = \int_0^\infty \kappa_V(r)\,\txd r.
\end{equation}
If we use the triplet $(\tau_V,\alpha,X_\lambda)$ as parameters,
the opacity at an arbitrary wavelength $\lambda$ can be written as
\begin{equation}
    \kappa_\lambda(r)
    =
    \frac{2}{\sqrt{\pi}}\,
    \frac{\Gamma\left(\frac{\alpha}{2}\right)}
    {\Gamma\left(\frac{\alpha-1}{2}\right)}\,
    \frac{X_\lambda\tau_V}{c}\,
    \left(1+\frac{r^2}{c^2}\right)^{-\alpha/2}.
\label{kappa}
\end{equation}

\subsubsection{The phase function}
\label{phafu.sec}

We can safely assume that the phase function does not depend
independently on the four variables $(\bfk,\bfk')$, but only on
the angle between the two directions $\bfk$ and $\bfk'$. Setting
$\bfk\cdot\bfk' = \cos\alpha$, we can write the phase function as
$\Phi(\cos\alpha)$. Probably the most widely adopted phase
function that can describe anisotropic scattering is the one named
after Henyey \& Greenstein (1941). It has been derived
empirically, as a description of the scattering of light in
reflection nebulae in the Galaxy, and has a simple analytical
form,
\begin{equation}
    \Phi(\cos\alpha)
    =
    \frac{1-g^2}{(1+g^2-2g\cos\alpha)^{3/2}}.
\label{hg}
\end{equation}
This family of phase functions contains the so-called asymmetry
parameter $g$, which is a measure for the anisotropy of the
scattering. In particular, for $g=0$ the scattering is isotropic,
whereas for $g=1$ the scattering is completely forward.

\subsubsection{The optical properties}
\label{optprop.sec}

\begin{table}
\centering \caption{The adopted optical properties of the dust
grains, at the central wavelengths of the standard optical and
near-infrared wavebands. Tabulated are the relative extinction
coefficient $X_\lambda$, the scattering albedo $\omega_\lambda$
and the asymmetry parameter $g_\lambda$.} \label{dustprop.tab}
\begin{tabular}{ccccc} \hline
band & $\lambda$\,(nm) & $X_\lambda$ & $\omega_\lambda$ & $g_\lambda$ \\
\hline
$U$ & 360 & 1.52 & 0.63 & 0.65 \\
$B$ & 440 & 1.32 & 0.61 & 0.63 \\
$V$ & 550 & 1.00 & 0.59 & 0.61 \\
$R$ & 700 & 0.76 & 0.57 & 0.57 \\
$I$ & 850 & 0.48 & 0.55 & 0.53 \\
$J$ & 1250 & 0.28 & 0.53 & 0.47 \\
$H$ & 1650 & 0.167 & 0.51 & 0.45 \\
$K$ & 2200 & 0.095 & 0.50 & 0.43 \\ \hline
\end{tabular}
\end{table}

From the previous subsections, we see that, in total, we need five
parameters to describe a dust component at a given wavelength
$\lambda$: the {\em V}-band optical depth $\tau_V$, the dust
exponent $\alpha$, the relative extinction coefficient
$X_\lambda$, the scattering albedo $\omega_\lambda$ and the
asymmetry parameter $g_\lambda$. In principle, all of these
parameters could be considered as free parameters. In order to
limit the degrees of freedom, however, we decided not to consider
the optical properties of the dust as a set of free parameters,
but to use a fixed set of dust parameters. The values we use are
listed in Table \ref{dustprop.tab}. They are taken from Gordon,
Calzetti \& Witt (1997), who derived them from an interpolation
between a large set of empirical data of typical Milky Way dust.
Notice in particular that the scattering albedo does not vary
greatly within the optical and near-infrared regime. It is always
slightly greater than 0.5, i.e.\ the probability for absorption is
always slightly smaller than the probability for scattering.

\subsection{A template model}
\label{templ.sec}

Our models contain three parameters, the orbital structure
parameter $q$, the $V$-band optical depth $\tau_V$ and the dust
exponent $\alpha$. By varying these, we can describe a set of
dusty elliptical galaxy models with a large variety in internal
dynamics, dust content and star-dust geometry. We apply the same
strategy as in Paper~I to present the results of our modelling.
First, we keep the geometry of the dust component fixed, and we
investigate how dust distributions with various optical depths
affect the light profile and observed kinematics of our three
Plummer models. Next, we keep the optical depth fixed and consider
a set of star-dust geometries by varying the dust exponent
$\alpha$.

We therefore require a template model, where we can vary one
parameter while keeping the other one fixed. For the optical depth
of this template model, we choose $\tau_V=1$, a value that could
be in agreement with both the far-infrared emission and the colour
gradients in elliptical galaxies (Witt et al.\ 1992; Goudfrooij \&
de Jong 1995; Wise \& Silva 1996). For the dust geometry of the
template model, we adopt a modified Hubble profile, characterized
by setting $\alpha=3$. It is shallower than the stellar
distribution, which seems indicated by colour gradient models
(Wise \& Silva 1996). As a note we would like to stress that both
the optical depth and geometry of the diffuse dust component in
elliptical galaxies are very poorly known, apart from the solid
result that FIR emission indicates that it is present. We return
to this issue in Section \ref{sec.detdustdist}.

Finally, we need to choose a template wavelength to present our
results. Because the present paper focuses on the effects of dust
attenuation on the observed kinematics, and because most of the
kinematical observations are presently conducted at optical
wavelengths, we will work in the $V$ band, unless mentioned
otherwise.

\section{Results}
\label{results.sec}

\subsection{Putting the method to the test}
\label{test.sec}

\begin{figure*}
\centering
\includegraphics[clip,width=\textwidth]{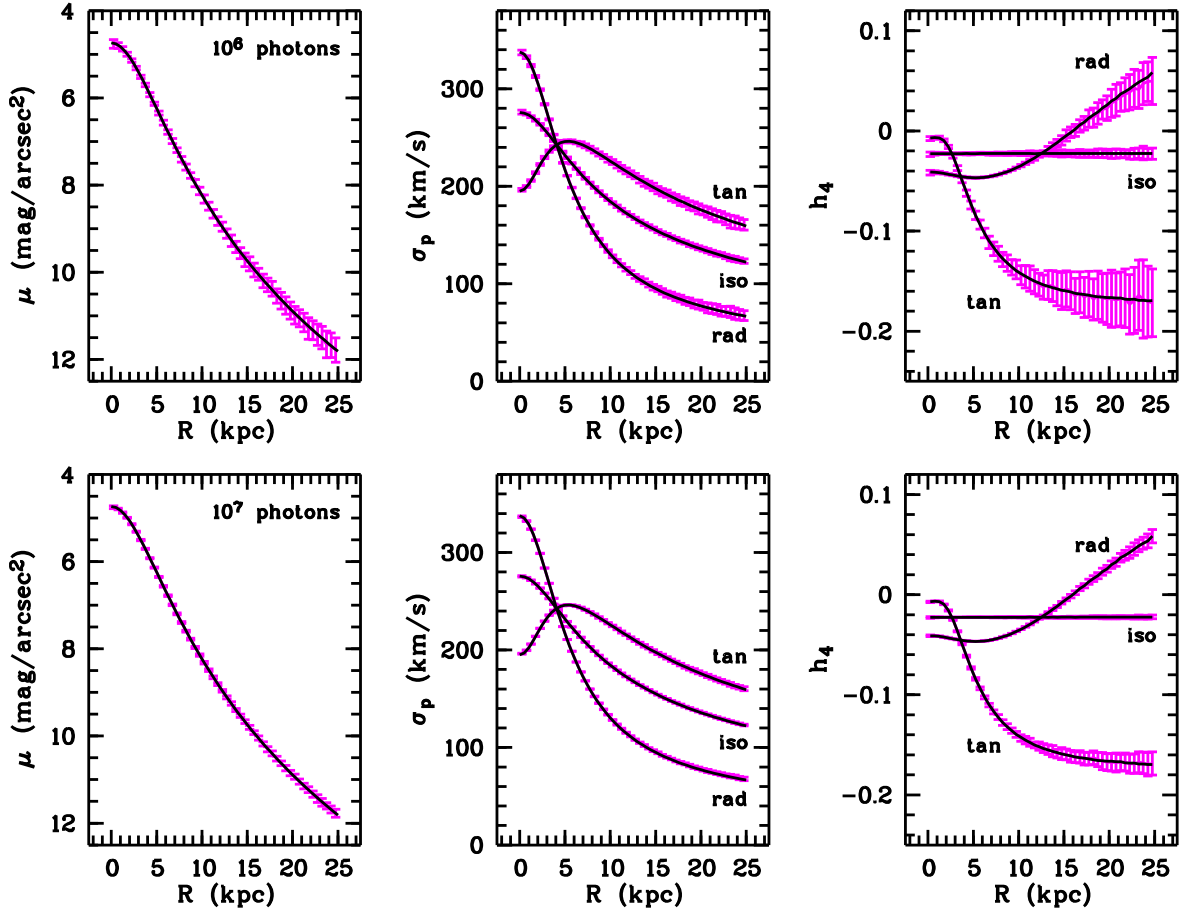}
\caption{A comparison of the results of the Monte Carlo code in
\nodu\ mode (grey curves with error bars) and the corresponding
analytical results (black lines). The upper and lower panels
correspond to different values for the total number of emitted
photons. Shown are the surface brightness profile, the projected
velocity dispersion and the $h_4$ profile, for the three different
orbital modes considered.}
\label{err2.eps}
\end{figure*}

The Monte Carlo routine yields results in three modes, the \nodu,
\abso\ and \dust\ modes. Whereas we are of course mainly
interested in the \dust\ results, the other data sets are useful
as a check for the accuracy of our results. If dust is not taken
into account, the light profile and the projected kinematics can
be calculated analytically for the galaxy models we consider
(Dejonghe 1987). If only dust absorption is taken into account,
they can be calculated through one single quadrature (Paper~I). In
Fig.\ \ref{err2.eps} we compare the results of our Monte Carlo
code with the corresponding analytical results, for two different
values of $N$, the total number of emitted photons. Even for $N$
as low as $10^6$, the minimum number of photons we consider, the
analytical results are very well reproduced, and are everywhere
within the error bars.

\subsection{The light profile}
\label{lightprofile.sec}

\subsubsection{Dependence on the optical depth}

\begin{figure*}
\centering
\includegraphics[clip,width=\textwidth]{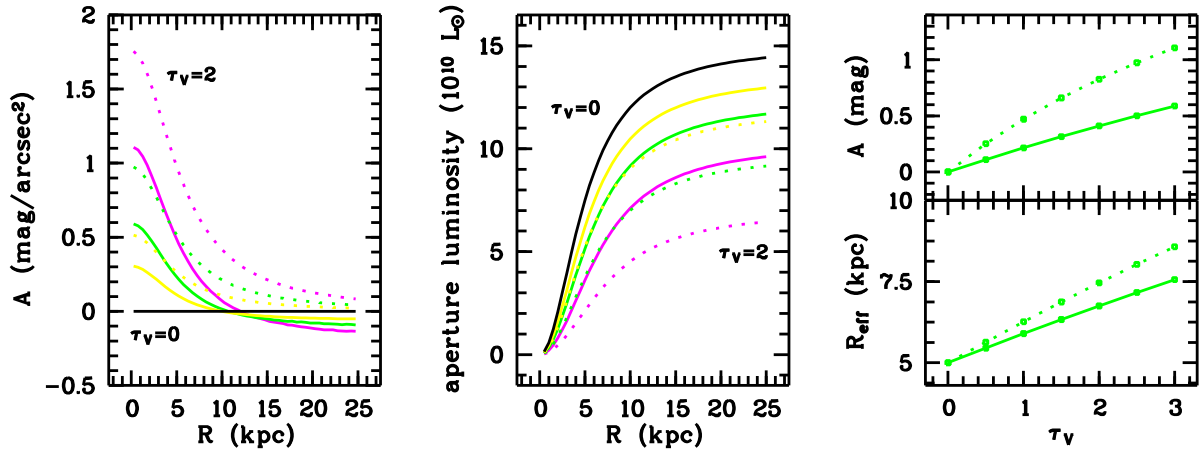}
\caption{The effects of dust attenuation on the light profile, as
a function of the optical depth. The left panel shows the
attenuation profile of the galaxy, i.e. the fraction of the light
output attenuated by the dust in magnitude units. It is shown for
absorption only (dashed lines), and for full attenuation (solid
lines), for optical depths $\tau_V=0, \tfrac{1}{2}, 1$ and $2$ --
only the extreme ones are labelled. The middle plot shows the
cumulative luminosity function, i.e.\ the fraction of the total
emitted luminosity detected inside an aperture of radius $R$. The
plots on the right-hand side show the effect of dust attenuation
on the total attenuation and the effective radius, as a function
of optical depth.}
\label{lipr.eps}
\end{figure*}

In Fig.\ \ref{lipr.eps} we demonstrate how the light profile is
affected when both absorption and scattering are included. In the
center of the galaxy (where the lines of sight contain most of the
dust), the attenuation is strongest, and the attenuation decreases
as one goes to the outer regions. As a consequence, also the
apparent luminosity decreases, and the apparent size of the core
(as measured by the effective radius $R_e$) increases as a
function of $\tau_V$.

Within the first few kpc, dust attenuation has at first order the
same effect on the light profile as absorption alone. Roughly, the
effects of a attenuating dust component with optical depth
$\tau_V$ can be approximated by a purely absorbing dust component
with effective optical depth $\tau_V/2$. This approximation has a
natural explanation: it follows from assuming that the scattering
is purely forward. Indeed, the phase function corresponding to
forward scattering is a simple Dirac delta function,
$\Phi(\bfk,\bfk')=4\pi\,\delta(\bfk-\bfk')$, and substituting this
into the RTE (\ref{RTE_dust}), we find
\begin{equation}
    \frac{\txd I}{\txd s}(\bfr,\bfk)
    =
    \ell(\bfr) - (1-\omega)\,\kappa(\bfr)\,I(\bfr,\bfk).
\label{RTE_forw}
\end{equation}
This radiative transfer problem is completely analogous to a
radiative transfer problem where only absorption is accounted for,
but where the optical depth of the dust component is diminished
with a factor $1-\omega\approx\tfrac{1}{2}$. This approximation is
an appealing way to estimate the effects of scattering without
elaborate and costly radiative transfer calculations. An argument
in favour of this approximation is that multiple scattering events
tend to wash out the effects of anisotropy of the scattering phase
function. Hence, in media with a large opacity, any phase function
can in principle be adopted, including the degenerate one
corresponding to forward scattering (e.g.\ Di Bartolomeo, Barbaro
\& Perinotto 1995).

However, the effects of scattering should never be underestimated.
In a previous study on the RTE in plane-parallel geometry, we
investigated several ways often used in the literature to
approximate scattering (Baes \& Dejonghe 2001b). One of our main
conclusions was that none of these methods provides a satisfactory
approximation to the exact solution. The reason is that the
physical process of scattering has a completely different nature
than absorption, because it changes the path along which the
photons propagate. The effect of scattering is that photons have a
preferential direction in which to leave the galaxy. In a
plane-parallel geometry, photons prefer the face-on direction to
leave the galaxy, because when a photon is scattered into that
direction, its chances to leave the galaxy are larger (the optical
depth along the path is shorter) than when it is scattered into
inclined directions. An analogous reasoning holds for spherical
geometry, where photons will generally prefer large projected
radii to leave the galaxy.\footnote{At any position in a spherical
galaxy, the optical depth is obviously smallest along the path in
the radial direction, i.e.\ the path directed away from the center
of the galaxy. One would therefore be inclined to think that the
photons will prefer to leave the galaxy through the central lines
of sight. However, the connection between lines of sight and
directions is not one-to-one: to every possible line of sight a
photon can be scattered into (i.e.\ to every $R\leq r$),
correspond two directions, one directed towards the center of the
galaxy and one towards the edge of the galaxy. The probability
that a photon, scattered at $r$ into a line-of-sight $R$, will
leave the galaxy, equals the weight function ${\mathcal{K}}(R,r)$,
defined in Section~2 of Paper~I. For fixed values of $r$, the
function ${\mathcal{K}}(R,r)$ is an increasing function of $R$,
hence having its maximal value at $R=r$. Photons will hence on
average more easily leave the galaxy at large projected radii.} As
a consequence, the net effect of scattering is that photons are
scattered out of lines of sight close to the center of the galaxy,
into lines of sight with a large projected radius. In the outer
regions of the galaxy, these extra photons will reduce the loss of
radiation due to absorption. At very large projected radii, the
attenuation will even be negative, i.e.\ the galaxy will even
appear brighter than when dust extinction is not taken into
account (Fig.\ \ref{lipr.eps}). These results are in agreement
with those found by Wise \& Silva (1996), and they will be very
important when we investigate the full effects of attenuation on
the observed kinematics.

\subsubsection{Dependence on the dust geometry}
\label{lipr_dustgeom}

\begin{figure*}
\includegraphics[clip,width=\textwidth]{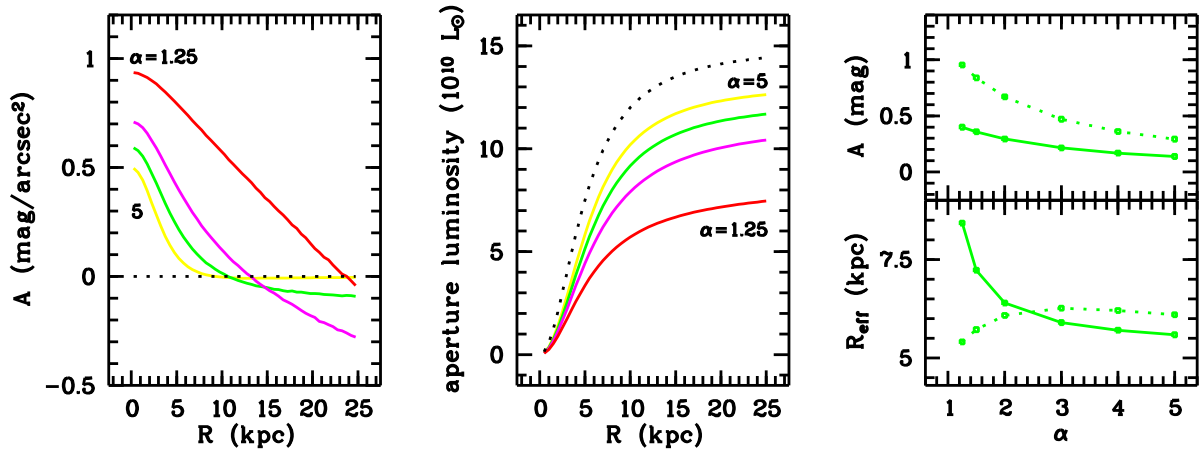}
\caption{The effects of dust attenuation on the light profile, as
a function of the dust geometry. The four panels in this figure
are analogous to those in Fig.\ \ref{lipr.eps}. The models shown
here all have $\tau_V=1$, but different values for $\alpha$. The
values shown are $\alpha=1.25$, 2$, 3$ and 5. The black dotted
lines in the left and middle panels correspond to the model
without dust attenuation. The grey dotted lines on the right-hand
side represent models with only absorption taken into account, as
in Fig.\ \ref{lipr.eps}.}
\label{lipralp.eps}
\end{figure*}

In Fig.\ \ref{lipralp.eps} we show the effects of varying the dust
exponent on the light profile of our galaxy models. In the central
regions of the galaxy, the combined effects of scattering and
absorption on the light profile can roughly be approximated by
pure absorption with an effective optical depth
$\tau_V^{\text{eff}}=(1-\omega)\,\tau_V$. Therefore, the
dependence on the dust exponent are similar to those when
absorption only is taken in to account (Paper~I): the attenuation
is stronger for extended dust distributions ($\alpha$ small) than
for centrally concentrated ones ($\alpha$ large). Because the
central regions emit most of the light and therefore dominate the
total observed luminosity, also the total attenuation will
decrease as a function of $\alpha$.

At large projected radii, the dust has another effect on the light
profile: the net effect is that photons scattered into these lines
of sight cause a negative attenuation, i.e.\ the galaxy appears
brighter. This effect should be stronger for extended dust
distributions, because these distributions imply more dust in the
outer regions of the galaxy, and hence an enhanced probability to
be scattered into the outer lines of sight. This can indeed be
observed in Fig.\ \ref{lipralp.eps}. In particular, this negative
attenuation is nearly non-existent for a model where dust and
stars have the same spatial distribution ($\alpha=5$), whereas it
is clearly noticeable if the dust distribution is shallower than
the stellar distribution.

\subsection{The observed kinematics}

\subsubsection{Dependence on the optical depth}

In an optically thin galaxy, the LOSVD is formed by summing the
contribution of the line of sight velocities of all stars that are
situated along that line of sight. If absorption is included in
the projection process, still the same stars on the line of sight
contribute to the LOSVD, but the contribution of each star is
weighted by the amount of starlight that is able to survive the
absorption and reach the observer. The net effect is that, for a
given line of sight, the stars at the outer parts of the line of
sight contribute relatively more to the LOSVD than the stars in
the central parts (Paper~I). Because the largest line-of-sight
velocities along a given line of sight are usually found around
the tangent point, the LOSVDs will be biased towards smaller
velocities. In particular, the projected velocity dispersion
generally decreases if absorption is taken into account. Only for
the outer lines of sight of galaxies with a very radially
anisotropic orbital structure, the largest line-of-sight
velocities are found in the outer regions. Indeed, the stars at
the tangent point have small line-of-sight velocities, because
their (radial) orbits are nearly perpendicular to the line of
sight. We showed in Paper~I that the effects of dust absorption on
the LOSVDs are only considerable for large optical depths
($\tau_V>5$), which are probably not appropriate for elliptical
galaxies.

If scattering is taken into account, the situation changes
drastically. The net effect of scattering, at least for the light
profile, is that photons are taken away from the central lines of
sight and sent into lines of sight at larger projected radii. Now
consider a photon that is emitted by a star near the center of the
galaxy, and that, after one or several scattering events,
propagates towards the observer at a large projected radius.
Although this photon will contribute to the LOSVD at this large
projected radius, it carries the velocity information from the
emitting star. Notice that this star does not physically belong to
that line of sight. Hence, when scattering is taken into account,
the LOSVDs aren't LOSVDs anymore in the strict meaning of the
word: the LOSVD at a certain line of sight can contain information
of stars at totally different lines of sight. More generally,
every single star in the galaxy will contribute to every single
LOSVD, whereby its contribution will be weighed by the number of
photons that leave the galaxy along that line of sight.

Beside the photons that disappear from the line of sight due to
attenuation, we hence also have to account for the photons
scattered into the line of sight, which contribute the additional
kinematical information from stars that physically don't belong
there. How this process affects the LOSVDs is more complex than
the effects due to the photons taken away from the line of sight.
In particular, this effect is different for lines of sight that
pass through the center of the galaxy ($R\approx0$) and lines of
sight at larger projected radii ($R\gg c$). \\

\begin{figure*}
\centering
\includegraphics[clip,width=\textwidth]{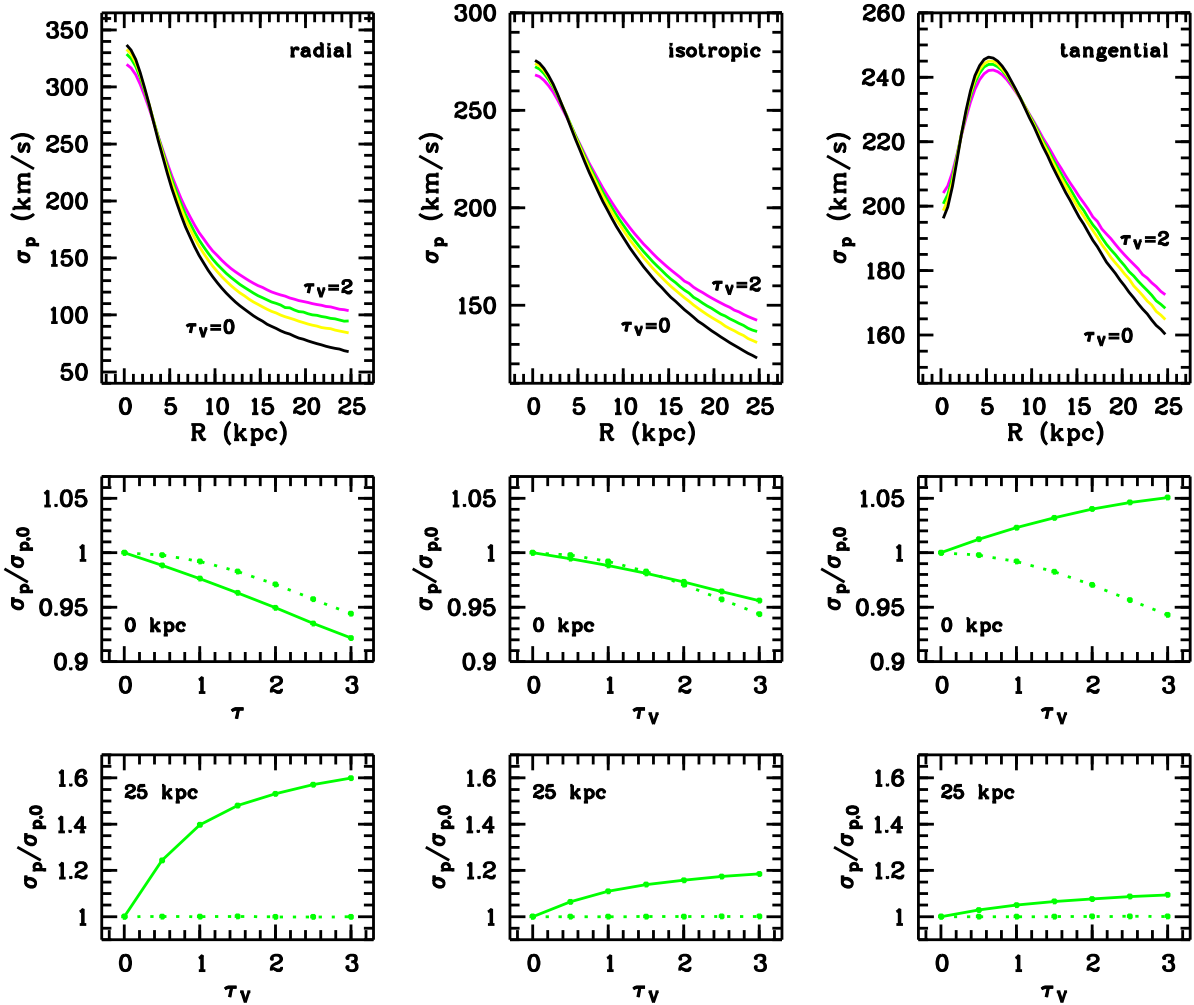}
\caption{The effects of dust attenuation on the observed velocity
dispersion, as a function of the optical depth. In the top panels,
the projected velocity dispersion profiles are plotted, for
different values of $\tau_V$. We have used the same values of
$\tau_V$ and the same layout as in Fig.\ \ref{lipr.eps}. The left,
middle and right columns correspond to our radial, isotropic and
tangential models respectively. The three panels on the middle row
show the central projected velocity dispersion as a function of
$\tau_V$. The values are normalized by $\sigma_{p,0}$, the central
projected velocity dispersion in an optically thin galaxy. The
solid lines represent the effect of dust attenuation, whereas in
dashed lines we show the effects of absorption only. Analogously,
the three bottom panels show the effects of dust attenuation,
respectively absorption, on the normalized projected velocity
dispersion at $R=25$\,kpc.} \label{disp.eps}
\end{figure*}

If scattering is not taken into account, the LOSVD at the central
line of sight $R=0$ only contains information on the radial
velocity component of stars, because along the central line of
sight $v_\parallel=v_r$. Because the vast majority of the stars
along this line of sight reside in the central regions, the
central LOSVDs will be dominated by the distribution of radial
velocities in the galaxy center. If scattering is included, the
central line of sight will also contain photons scattered into it,
that would normally leave the galaxy at a larger projected radius.
The vast majority of these photons will also originate from near
the central region, but the line-of-sight velocities carried by
them are not necessarily the radial velocity components of the
stars that emitted them. These photons contaminate the LOSVD with
tangential velocities: instead of containing pure radial velocity
information, it will reflect a mix of radial and tangential
components. As a consequence, the effect of scattering will depend
on the ratio of radial to tangential velocity components in the
central regions of the galaxy, i.e.\ on the orbital structure. In
the panels on the middle row in Fig.\ \ref{disp.eps}, we plot the
effect of increasing optical depth on the central projected
velocity dispersion.

For isotropic galaxies, radial and tangential velocities are in
balance throughout the galaxy, such that the photons scattered
into the central lines of sight will hardly affect the LOSVD. The
total effect of attenuation on the central LOSVDs will be
dominated by the absorption effect, i.e.\ a bias towards smaller
line-of-sight velocities. In particular, the central projected
dispersion decreases if dust is taken into account, in a very
similar way as when only absorption is taken into account.

\begin{figure*}
\centering
\includegraphics[clip,width=\textwidth]{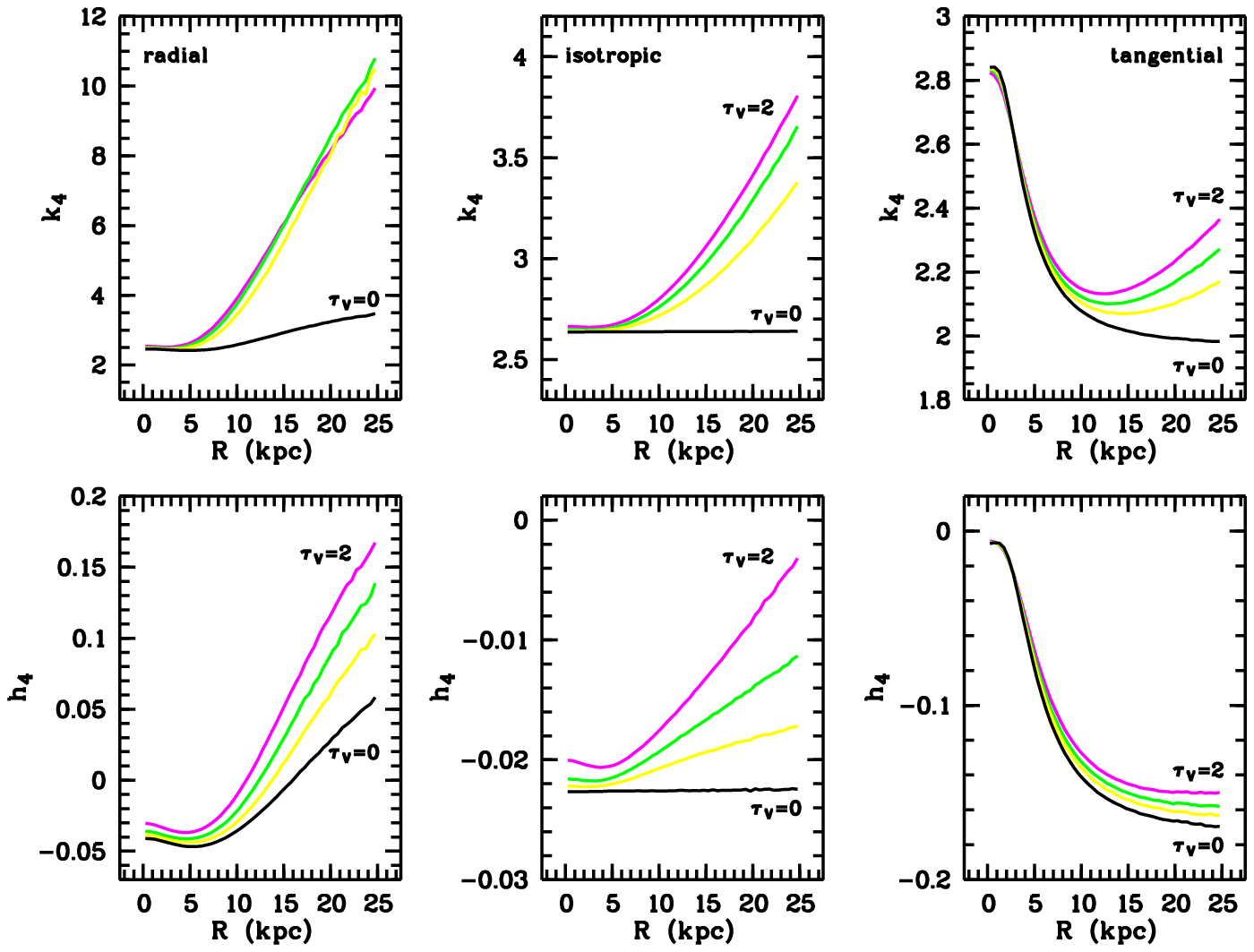}
\caption{The effects of dust attenuation on the observed LOSVD
shape, parametrized by the kurtosis $k_4$ and Gauss-Hermite $h_4$
coefficient, as a function of the optical depth. As in Fig.\
\ref{disp.eps}, the left, middle and right columns correspond to
the radial, isotropic and tangential models respectively. The
adopted optical depths and layout are similar to those used in
Fig.\ \ref{lipr.eps}.} \label{k4h4.eps}
\end{figure*}

\begin{figure*}
\centering
\includegraphics[clip,width=\textwidth]{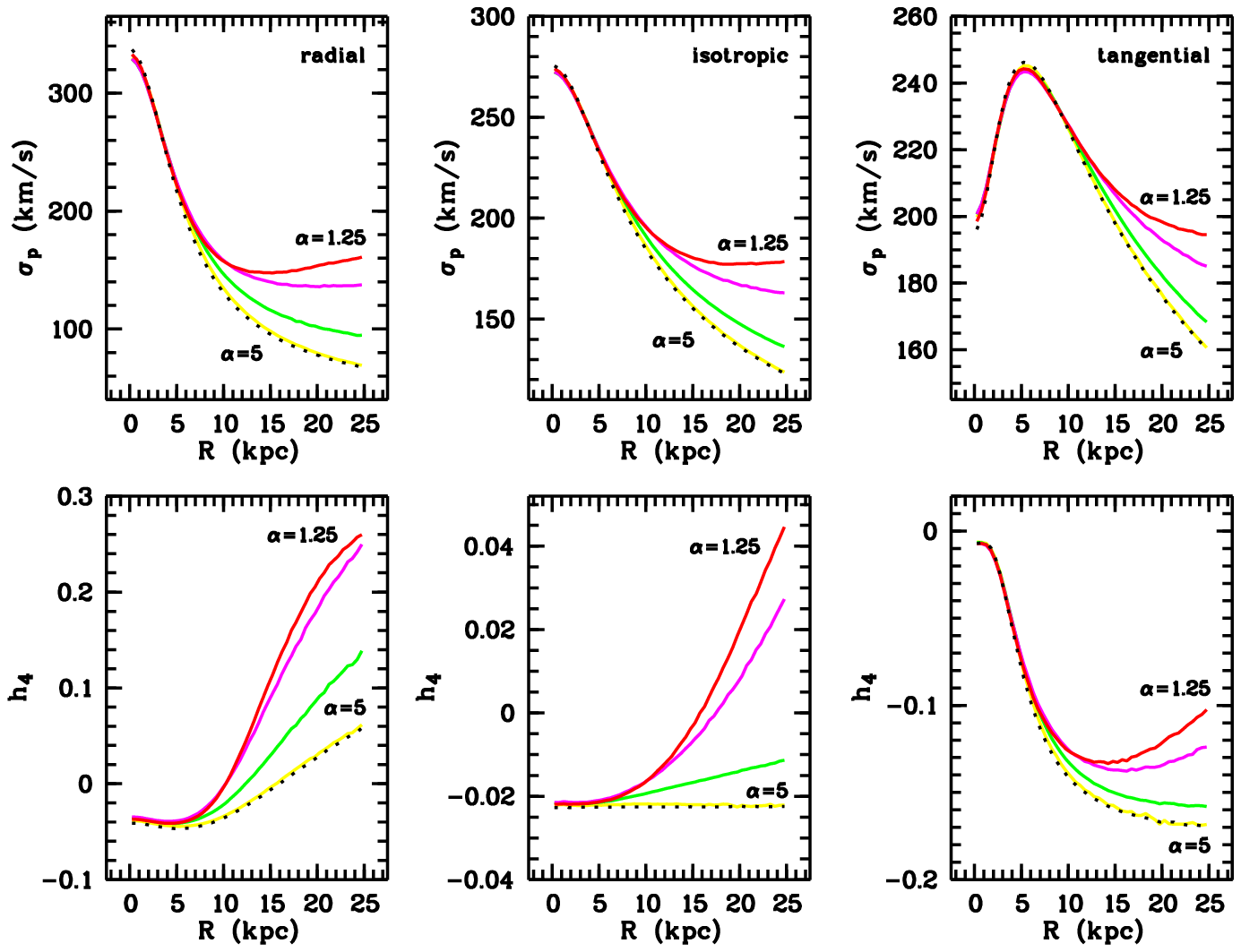}
\caption{The effects of dust attenuation on the observed
kinematics, as a function of the dust geometry. Shown are the
projected dispersion profile and the $h_4$ profile, for different
values of the dust exponent $\alpha$. The presented models and the
layout are the same as in Fig.\ \ref{lipralp.eps}. As in the
previous figures, the three column correspond to the three
different Plummer models.} \label{alpkin.eps}
\end{figure*}

In a radially anisotropic galaxy, stars have on average a larger
radial than tangential velocity component. Stars scattered into
the central line of sight that contribute part of their tangential
velocity, will therefore bias the LOSVD towards smaller
velocities. Because the effect of absorption is also a bias
towards smaller velocities, the total effect of attenuation on the
central LOSVDs of radially anisotropic galaxies is to turn them
more peaked, i.e.\ to decrease the central projected dispersion.
The strength of the effect is stronger than the effect of
absorption alone.

On the contrary, for tangentially anisotropic galaxies, the
tangential velocity of the stars generally exceeds their radial
velocity. As a consequence, if the central LOSVD is contaminated
with tangential velocity information, it will be biased towards
larger line-of-sight velocities. Two processes hence affect the
LOSVD in an opposite way: absorption (and scattering of photons
out of the line of sight) favors smaller velocities, whereas
scattering of photons into the line of sight biases the LOSVD
towards larger velocities. The way the central LOSVDs are affected
depends on which of both mechanisms is stronger. For the modest
optical depths appropriate in elliptical galaxies, the scattering
effect dominates, and the central projected dispersion increases
due to scattering. \\

For lines of sight at large projected radii, the situation is
completely different. We argued that the net effect of scattering
is that photons are scattered from the central lines of sight into
the outer lines of sight. These scattered photons will strongly
affect the observed kinematics. Indeed, the majority of these
scattered photons are emitted by stars in the central regions of
the galaxy. They carry along the kinematical information of the
stars that emitted them, i.e.\ the typical velocities appropriate
in the central regions of the galaxy. These are on average larger
than the typical line-of-sight velocities in the outer regions of
the galaxy, where the kinetic energy of the stars is much lower.
As a consequence, they will contaminate the LOSVDs with high
velocities. In particular, it should be noted that lines-of-sight
velocities will be observed in the LOSVDs, that would normally be
impossible at these projected radii. If scattering is not taken
into account, the maximal line-of-sight velocity of the LOSVD at a
projected radius $R$ is the escape velocity
$v_{\text{esc}}(R)=\sqrt{2\psi(R)}$, where $\psi(R)$ is the
potential at the tangent point $r=R$. If scattering is taken into
account, the photons observed at the line-of-sight $R$ can
originate from stars in the center of the galaxy where the
line-of-sight velocity can be larger that the escape velocity at
$R$. These large line-of-sight velocities will cause "forbidden"
high-velocity wings in the outer LOSVDs. As a result, the
projected dispersion at large projected radii will increase.

In the top panels of Fig.\ \ref{disp.eps}, we plot the projected
velocity dispersion profiles for various values of the optical
depth. Clearly, the velocity dispersion increases significantly in
the outer regions. The strength of this increase depends on the
orbital structure of the galaxy. Radially anisotropic models are
more strongly affected than tangential ones, because their
(optically thin) outer LOSVDs are more strongly peaked, and thus
more vulnerable to the contribution of photons from high-velocity
stars in the center. This is clearly shown in the bottom row of
Fig.\ \ref{disp.eps}, where the effect of dust attenuation on the
projected velocity dispersion at large projected radii is shown as
a function of the optical depth. The effects are significant: for
example, for an optical depth of unity, the projected dispersion
increases with more than 40 per cent for the
radial model. \\

Of course, the LOSVDs are not completely determined by the
projected velocity dispersion alone, which just gives a measure
for the broadness of the LOSVD. The extra information contained in
the (symmetric) LOSVDs, the actual shape of the LOSVDs, can be
represented by either the kurtosis or the $h_4$ parameter. In the
top panels of Fig.\ \ref{k4h4.eps} we show how dust attenuation
affects the shape of the LOSVDs of our Plummer galaxies, as
quantified by the kurtosis. In the inner regions, the effect is
fairly small. As the effect on the projected dispersion, its sign
is dependent on the orbital structure: for the radial and
isotropic models, the kurtosis increases with optical depth,
whereas it decreases for the tangential model. At large projected
radii, the kurtosis increases spectacularly if dust attenuation is
taken into account, and again, the effect is much stronger for
radial than for tangential models. The reason for this increase
are the high-velocity wings in the LOSVDs.

The bottom panels in Fig.\ \ref{k4h4.eps} show the effects of dust
attenuation on the shape of the LOSVDs of our models, as
quantified by the $h_4$ parameter. The effects are comparable to
those on the kurtosis. This comes as no surprise, because kurtosis
and $h_4$ are proportional to first order (van der Marel \& Franx
1993).

\subsubsection{Dependence on the dust geometry}

In Section \ref{lipr_dustgeom}, we showed that the contribution of
scattered photons to the surface brightness at large radii depends
rather critically on the dust geometry. If, on the one hand, the
dust distribution is shallower than the stellar distribution, the
contribution of these photons is important, and they even cause a
negative attenuation at the outer lines of sight. If, on the other
hand, the dust follows the stellar distribution, the effects of
scattering are nearly negligible. Because we showed that these
scattered photons affect the observed kinematics rather strongly,
it can be expected that the effect of dust attenuation on the
observed kinematics will also be very sensitive to the dust
geometry.

In Fig.\ \ref{alpkin.eps} we illustrate how the observed
kinematics depend on the dust geometry. We find indeed that the
kinematics are much more affected for extended dust distribution
than for centrally concentrated ones. In particular, when the dust
has the same spatial distribution as the stars, the effects of
scattering on the observed kinematics are nearly negligible. On
the contrary, when the dust density decreases very slowly, the
outer regions have a large dust-to-stars ratio, such that the
photons scattered into these lines of sight form a large fraction
of the total number of photons that contribute to the LOSVDs. The
high-velocity stars gradually contribute more as the dust geometry
becomes more extended. As a consequence, the projected velocity
dispersion profile increases significantly with decreasing
$\alpha$ at large projected radii, and also the effect on the
$h_4$ shape parameter depends strongly on the dust geometry. The
dust distribution is clearly an important parameter in our models.

\subsubsection{The influence of optical property gradients}

\begin{figure*}
\centering
\includegraphics[clip,width=\textwidth]{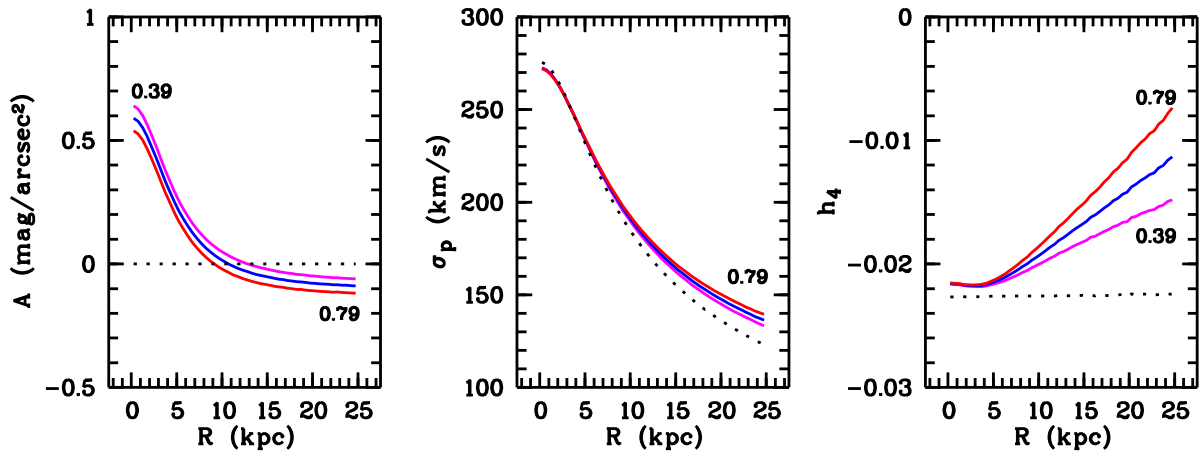}
\caption{The effect of a dust albedo gradient on the attenuation
curve and the observed kinematics, more precisely the projected
dispersion and $h_4$ profiles. The adopted model is the isotropic
template model in the $V$ band, but with a variable dust albedo of
the form (\ref{albgr}), with $\omega_0=\omega_V=0.59$, and three
different values of $\omega_\infty$, being 0.79, 0.59 and 0.39.
The black dotted curves correspond to the model without dust
attenuation.} \label{albgrad.eps}
\end{figure*}

From the previously obtained results, we know that the effects of
dust attenuation on the observed kinematics are due to photons
emitted by high-velocity stars in the centre of the galaxy,
scattered in the outer regions on lines of sight at large
projected radii. The strength of these effects will hence depend
on the probability that a photon from a high velocity star will be
scattered onto outer lines of sight. There are various factors
that can contribute to this probability. We already encountered
two of them: the larger the dust content of the galaxy (high
$\tau$) and/or the more extended the dust distributions (small
$\alpha$), the greater the probability for scattering at large
radii, and hence the stronger the effects on the observed
kinematics. But also the optical properties of the dust can
contribute to the number of scattering events that send photons
from the inner lines of sight to outer lines of sight. Imagine,
for example, that the scattering would be completely forward at
large radii, or that the scattering albedo would become negligible
in the outskirts of the galaxy. In either case, the probability
that a photon emitted in the centre would reach the observer along
a line of sight at large projected radii would be very small, such
that the observed kinematics would hardly be affected by dust
attenuation.

So far, we have adopted the assumption that the optical properties
of the dust are equal all over the galaxy. However, this
assumption might not always be satisfied in real elliptical
galaxies. Indeed, gradients in e.g.\ the metallicity, the stellar
radiation field or the X-ray luminosity density of a galaxy can
cause systematic changes in the size distribution and/or chemical
composition of the dust grains, which can result in gradients in
their optical properties. In our own Galaxy, the physical
properties of interstellar dust have been found to vary
substantially in different environments (Witt, Bohlin \& Stecher
1984; Mathis \& Cardelli 1992).

With our Monte Carlo technique, we can easily include optical
properties gradients in our models. The strongest effect can be
expected for a gradient in the scattering albedo, because
$\omega(r)$ is directly related to the number of scattering events
at a distance $r$. We add to the template model of Section
\ref{templ.sec} an albedo gradient of the form
\begin{equation}
    \omega(r)
    =
    \frac{\omega_0c^2+\omega_\infty r^2}{c^2+r^2},
\label{albgr}
\end{equation}
which changes smoothly from $\omega_0$ in the centre to
$\omega_\infty$ at large radii. In Fig.\ \ref{albgrad.eps} we show
the effect of such a gradient on the $\sigma_p$ and $h_4$
profiles, with the central albedo $\omega_0$ the value from Tab.\
\ref{dustprop.tab}, and a number of different values for
$\omega_\infty$. This figure demonstrates that the attenuation
effect on the observed kinematics becomes stronger for larger
values of $\omega_\infty$, which agrees with the prediction that
more scattering events correspond to a larger effect on the
observed kinematics. Notice that for the reasonably large
gradients in $\omega$ (more than 30 per cent), the differences
between the various kinematical profiles are relatively small, and
definitely within the observational errors.

\section{Discussion}
\label{discussion.sec}

\subsection{Dark matter halos around elliptical galaxies}

In the outer regions of the galaxy, the observed kinematics are
strongly affected by photons emitted in the central regions of the
galaxy, that leave the galaxy after one or more scattering events
along lines of sight with a large projected radius. Because these
photons are emitted by stars that generally have larger velocities
than the typical line-of-sight velocities appropriate along these
lines of sight, they bias the LOSVDs towards larger velocities,
and cause high-velocity wings. As a result, the projected velocity
dispersion and the shape parameters $k_4$ and $h_4$ increase
significantly at large projected radii.

These results are particularly important for the interpretation of
the stellar kinematical evidence for dark matter halos around
elliptical galaxies. For disc galaxies, the observational evidence
for the existence of dark matter halos is convincing: the H{\sc i}
rotation curves that remain flat or even rising out to very large
radii, provide a clear proof of their existence (Freeman 1970;
Faber \& Gallagher 1979). For elliptical galaxies, this important
tracer can generally not be used. A number of early-type galaxies,
most of them classified as S0s, have neutral or ionized gas discs
which can be used to estimate their dark matter content (Bertola
et al.\ 1993; Franx et al.\ 1994), but these galaxies are
exceptional cases and may not be representative for the general
class of elliptical galaxies. The most convincing evidence for the
existence of dark matter halos comes from measurements of the
density and temperature of their hot X-ray emitting atmospheres
(Forman, Jones \& Tucker 1985; Matsushita et al.\ 1998;
Loewenstein \& White 1999), and from gravitational lensing
(Griffiths et al.\ 1996; Keeton, Kochanek \& Falco 1998). This
evidence indicates that elliptical galaxies must have very large
mass-to-light ratios at large radii, but, unfortunately, they do
not contain much information about the detailed structure of a
dark matter halo and its coupling to the luminous matter.

The most important way to trace dark matter halos around
elliptical galaxies at small scales is by studying the stellar
kinematics. With the present 8m class telescopes, stellar
kinematics can be reliably traced out to several effective radii.
If a dark matter halo is present, one expects the velocity
dispersion profile to drop only slowly or to remain constant with
projected radius. Such a behaviour was interpreted as a signature
for the presence of a dark matter halo (Saglia, Bertin \&
Stiavelli 1992; Saglia et al.\ 1993). However, a slowly decreasing
or nearly constant velocity dispersion profile can also be due to
a strong tangential anisotropy at large radii. This so-called
mass-anisotropy degeneracy (Gerhard 1993) can be broken by
studying the LOSVD shape parameters: galaxies with a tangentially
anisotropic orbital structure have a negative $h_4$. The
combination of a slowly decreasing velocity dispersion profile and
a positive $h_4$ at large projected radii, is generally
interpreted as a indication of the presence of a dark matter halo.
For a number of elliptical galaxies, the existence of dark halos
has been advocated by such evidence (Rix et al.\ 1998; Gerhard et
al.\ 1998; Kronawitter et al.\ 2000; Gerhard et al.\ 2001).

We show here that scattering off dust grains has the same effect
on the dispersion profile as a dark halo: the dispersion will
decrease more slowly than expected. Moreover, also the $h_4$
profiles are considerably biased towards larger values, such that
the signature of an intrinsically tangential anisotropy can be
weakened. Dust attenuation hence makes a foolproof detection of
dark matter halos from stellar kinematical evidence much more
complicated. It is obviously important to investigate to which
degree dust can reduce or eliminate the need for a dark matter
halo to explain the observed kinematics. Our first results
indicate that a dust component that is shallower than the stars
with $\tau_V=1$, has the same kinematic signature as a dark matter
halo that contains half of the total mass of the galaxy (Baes \&
Dejonghe 2001c). An in-depth investigation of this topic, however,
is beyond the scope of this paper, and a forthcoming paper will be
devoted to this problem.

\subsection{The determination of the dust distribution in
elliptical galaxies} \label{sec.detdustdist}

\begin{figure*}
\centering
\includegraphics[clip,width=\textwidth]{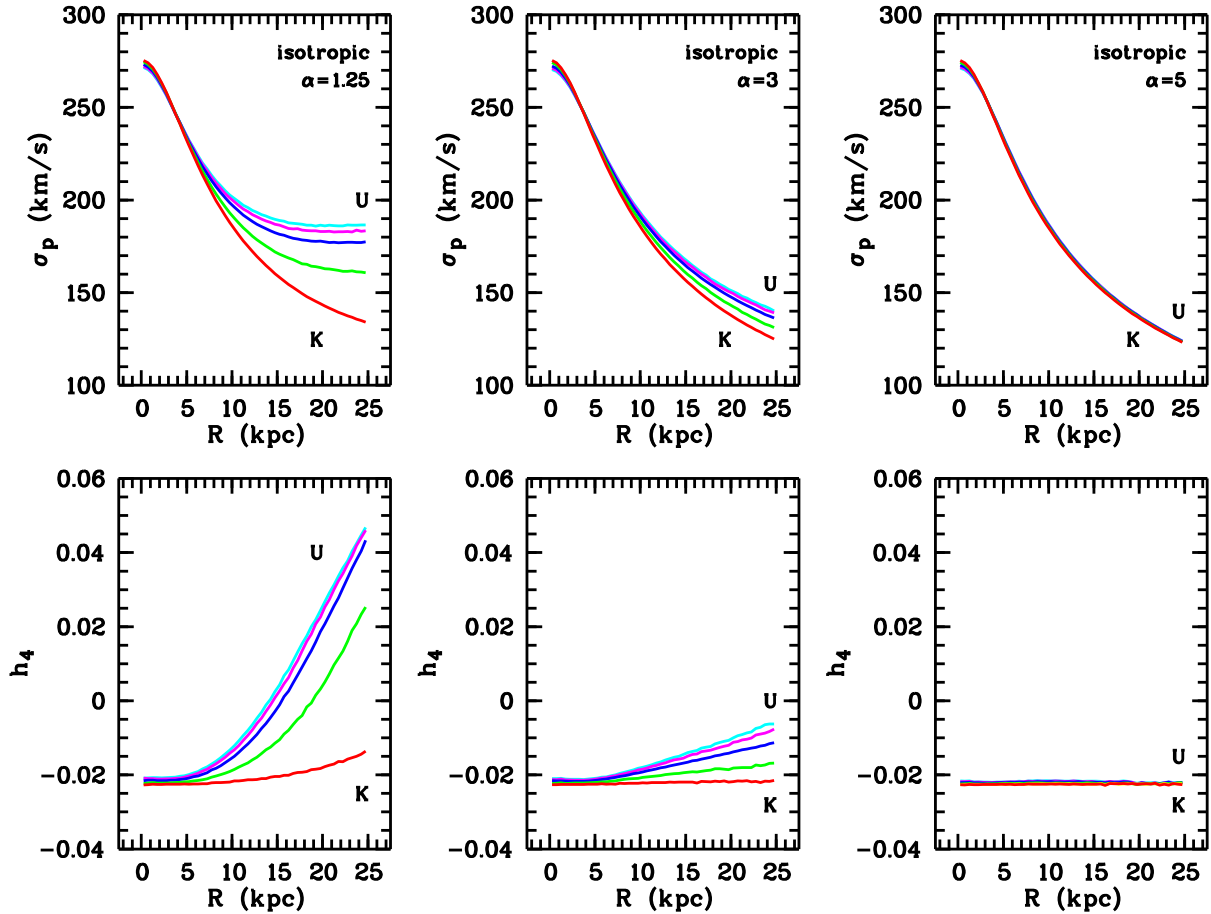}
\caption{The observed kinematics in different colours as a tool to
constrain the dust distribution in ellipticals. This figure shows
how the projected velocity dispersion profile (upper panels) and
the $h_4$ profile (lower panels) as they are observed in different
wavebands. In each panel, the wavebands shown are the {\em U},
{\em B}, {\em V}, {\em I}, {\em J} and {\em K} bands. All models
correspond to an isotropic Plummer model with a {\em V}-band
optical depth $\tau_V=1$, but the three columns correspond to
three different dust geometries: $\alpha=1.25$ (left), $\alpha=3$
(template model, middle) and $\alpha=5$ (right).}
\label{colorkin.eps}
\end{figure*}

In the previous subsection, dust attenuation is considered as
something troublesome or inconvenient -- as an obstacle that
prevents the observation of the true projected kinematics, and
therefore complicates the interpretation of the true dynamical
structure of elliptical galaxies. The fact that the observed
kinematics are seriously affected by dust attenuation, can also
serve in a more positive way: it can be used to trace the spatial
distribution of the diffuse dust component in elliptical galaxies.

This is illustrated in Fig.\ \ref{colorkin.eps}, where we
illustrate how the observed kinematics change with colour. The
$\sigma_p$ and $h_4$ profiles in blue bands are more strongly
affected by dust attenuation than those in red or near-infrared
bands, because the extinction efficiency decreases with wavelength
(Table \ref{dustprop.tab}). Combining this with the sensitive
dependence of the observed kinematics on the dust geometry, we see
that multi-colour kinematical profiles can help to constrain the
spatial distribution of the dust. Indeed, if the dust is shallower
than the stars ($\alpha$ small), the blue kinematics will be
significantly affected by dust attenuation, whereas this effect
will gradually become weaker if we look towards longer
wavelengths. If, on the other hand, the dust has a similar
distribution as the stars, the observed kinematics are hardly
affected, even in blue colours, such that we will see no change of
the kinematics with wavelength.

This method is also subject to a degeneration: differences between
the stellar kinematics at different wavelengths can also be due to
variations in the stellar populations, i.e.\ different absorption
lines can trace different stellar types which do not necessarily
need to be in the same dynamical state. The major problem,
however, is the observational challenge: the projected kinematics
need to be measured at different wavelengths with a sufficient
accuracy out to several effective radii. Whereas this is nowadays
possible in the optical, it is probably beyond the limit of the
possibilities of the current generation of telescopes and
instruments to do this at near-infrared wavelengths.

As a remark, we want to note that a similar principle has already
been applied in another context: a number of authors have tried to
constrain the dust content of disk galaxies by multi-wavelength
analysis of the apparent rotation curves. Bosma et al. (1992)
demonstrated how dust absorption can affect the observed rotation
curve of edge-on spiral galaxies, and argued that the comparison
of the optical and H{\sc i} rotation curves can be used as a new
opacity test. Applying this test to the edge-on spiral galaxy
NGC\,891, they found that at least the outer regions of this
galaxy must be optically thin. In a similar way, Prada et al.
(1994) measured the ionized gas rotation curve of the inclined
galaxy NGC\,2146 at different wavelengths. They found a few
discrepancies between the optical H$\alpha$ and the near-infrared
[S{\sc iii}] rotation curves in the center of the galaxy, which
they could attribute to a dust lane, and a nice agreement between
the rotation curves at large projected radii. Whence they
concluded that NGC\,2146 has to be largely transparent in its
outer regions.

\subsection{The central velocity dispersion in dusty galaxies}

In Paper~I, we found that the effects of absorption on the
observed kinematics are only noticeable in the most central
regions. For optical depths of order unity, absorption causes a
decrease of the central dispersion with a few per cent. When
scattering is taken into account, the effect of dust attenuation
on the central dispersion is more complicated to predict. In
particular, the nature (increase or decrease) and strength depend
on the orbital structure of the galaxy. Although the magnitude of
these effects is still modest (in particular compared to the
effects of dust attenuation at large projected radii), this has
some important implications.

A first area where the effects of dust attenuation should be taken
into account are mass estimates. Indeed, the observed central
velocity dispersion of galaxies is a parameter that is often
adopted to quantify the total mass of a galaxy. For elliptical
galaxies in the nearby universe, more accurate dynamical mass
estimates are available through modelling of the entire observed
kinematics. For other systems, however, in particular in the
high-$z$ universe, the central velocity dispersion is often the
only available kinematic property that can be measured with some
reliability. The effects of dust attenuation can hence lead to a
bias in simple mass estimates. Moreover, it is currently unclear
how much dust is present in high-$z$ galaxies -- a significant
amount of dust production can be expected at the early epochs of
star formation.

The central velocity dispersion also appears as an important
parameter in the discussion about the presence and masses of black
holes. It is believed that a significant fraction of the galaxies,
if not all, harbour a massive black hole in their inner regions.
The most reliable determination of the black hole masses in
quiescent galaxies is by means of spatially resolved kinematics.
This, however, represents a challenge for both observers and
modellers, because it requires a very high spatial resolution and
detailed dynamical modelling techniques. Mainly with
high-resolution {\em{HST}} data, black hole masses have been
determined for a number of nearby galaxies. It was found that they
are tightly correlated with the central velocity
dispersion\footnote{Gebhardt et al.\ (2000a) use an effective
dispersion in their relation, defined as the aperture dispersion
within 1\,$R_e$, whereas Ferrarese \& Merritt (2000) adopt an
aperture dispersion within an effective aperture of radius
$R_e/8$.} of the host galaxies by a relationship
$M_\bullet\propto\sigma^a$, with $a\approx4$ (Gebhardt et al.\
2000a; Ferrarese \& Merritt 2000). Black hole masses of nearby
AGNs, which can be determined by reverberation mapping (Blandford
\& McKee 1982; Kaspi et al.\ 2000), seem at first sight to satisfy
this relation as well, albeit with some more scatter (Gebhardt et
al.\ 2000b; Ferrarese et al.\ 2001). It is of course very tempting
to adopt this relation for the determination of black hole masses
in other galaxies, in particular at high $z$, where spatially
resolved kinematics are beyond the present observational
capabilities. However, because the lack of knowledge about the
dust content of these galaxies, some caution is advised when
applying this relation without taking attenuation effects into
account.

\subsection{Disk heating processes in spiral galaxies}

It is generally known that the velocity dispersion of stars in the
Galaxy increase with age (Wielen 1977). This is believed to be a
result of the gradual heating of a initially cold disk due to
irregularities in the gravitational potential. A number of
possible scattering agents have been proposed for this heating,
but the two most important mechanisms that contribute to this
heating are thought to be scattering off giant molecular clouds
(Spitzer \& Schwarzschild 1951, 1953) and spiral density waves
(Carlberg \& Sellwood~1985). The study of the velocity ellipsoid
in disk galaxies provides an interesting way to distinguish
between these mechanisms, because each of them leaves a different
kinematical signature. Spiral density waves are inefficient in
scattering the stars in the vertical direction, such that it will
result in a small $\sigma_z/\sigma_R$ ratio. On the other hand,
molecular clouds tend to scatter stars rather isotropically,
yielding an intermediate value for $\sigma_z/\sigma_R$. The axis
ratio of the velocity ellipsoid is hence a very useful tool to
identify the principal heating agent in spiral galaxies (Jenkins
\& Binney 1990; Merrifield, Gerssen \& Kuijken 2001).

Unfortunately, the determination of the shape of the velocity
ellipsoid in spirals is not straightforward. Gerssen, Kuijken \&
Merrifield (1997, 2000) showed that, in theory, it is possible to
constrain the shape of the velocity ellipsoid of intermediately
inclined spiral galaxies from the observed dispersion profiles on
the major and minor axes.

From the results in this paper, however, we anticipate that dust
attenuation has a strong effect on the projected dispersion
profiles in disk galaxies, because of two reasons. First, disk
galaxies contain large amounts of dust, typically several orders
of magnitude larger than the average elliptical galaxy. The
effects of attenuation are therefore likely too important to be
considered as a second-order effect. Second, the differences in
intrinsic velocity dispersion between bulge and disk are very
large. If a photon, emitted by a high-velocity bulge star,
propagates into the disk, and is scattered there such that it
leaves the galaxy at a large projected radius, it will contribute
a very large line-of-sight velocity to that LOSVD. A relatively
small number of bulge star photons can therefore already seriously
contaminate the LOSVDs and increase the observed projected
dispersion at these lines of sight. Moreover, both the dust
content and the bulge-disk ratio vary along the Hubble sequence,
which definitely complicates a simple picture. These ideas
definitely need a detailed investigation before major conclusions
can be drawn on the heating processes in disk galaxies from the
observed velocity ellipsoids.

\section{Conclusions}
\label{conclusions.sec}

The aim of this series of papers is to investigate the effects of
a diffuse dust component on the observed kinematics of elliptical
galaxies. We started in Paper~I by investigating the effects of
absorption only, neglecting the scattering effects. We found that,
for realistic optical depths, these effects are modest, i.e.\ of
the order of a few per cent in the central regions and completely
negligible at larger projected radii.

In this paper, we extended our models to include the effects of
scattering, which are usually considered as a second-order effect
compared to absorption. The underlying thought is that the main
effect of scattering is a reduction of the effects of absorption.
It is therefore believed that the effects of scattering can be
modelled by considering absorption with a reduced effective
optical depth. From this point of view, it is often considered not
worth the effort to properly include scattering in a proper way,
thereby bypassing a costly radiative transfer treatment.

If we would have adopted this prescription, there would have been
no reason to include scattering in our models. Indeed, because the
effects of absorption only on the observed kinematics are already
fairly small, the extra second-order effect of scattering should
be completely negligible. However, a number of authors, foremost
Witt et al.\ (1992), have shown that the effects of scattering are
important, even for small optical depths, and that any way of
neglecting or approximating them can lead to serious errors. We
confirmed this in a previous study on the effects of dust
attenuation on disk galaxies (Baes \& Dejonghe 2001b).

In this paper, we have clearly demonstrated that, concerning the
observed kinematics in elliptical galaxies, scattering can not be
considered as a second-order effect to absorption. On the
contrary, we find that the effects of dust attenuation are much
more complicated and fascinating if scattering is included in the
modelling. The way the kinematics of elliptical galaxies are
affected can differ drastically, depending on which line of sight
is considered, on the star-dust geometry and on the internal
orbital structure of the galaxy. The most striking effect is the
serious increase of both the velocity dispersion and the LOSVD
shape parameters at large projected radii, due to photons from
high-velocity stars scattered into the line of sight. This effect,
which complicates the interpretation of the stellar kinematical
evidence for dark matter halos around elliptical galaxies, has
absolutely no counterpart when only absorption is taken into
account. Results as these, which may seem unexpected at first
sight, can be easily understood, once the idea of scattering as a
second-order effect of absorption is abandoned.

\appendix
\section{The spatial LOSVD for the Plummer model}

\subsection{The spatial LOSVD}

The dynamical structure of a certain stellar population of a
galaxy is completely determined by its phase space distribution
function $F(\bfr,\bfv)$. Whereas this function gives us, at a
certain position in the galaxy, the three-dimensional distribution
of the velocities, it is interesting from an observational point
of view, interesting to know the (marginal) distribution of
velocities into an arbitrary direction $\bfk$, for example the
direction towards the observer. To calculate this distribution, we
construct a new cartesian reference system
$(\bfe_{\perp_1},\bfe_{\perp_2},\bfe_\parallel)$, such that
$\bfk=\bfe_\parallel$. The spatial LOSVD
$\phi(\bfr,\bfk,v_\parallel)$ is defined through the relation
\begin{equation}
    \ell(\bfr)\,\phi(\bfr,\bfk,v_\parallel)
    =
    {\mathcal{L}}\iint F(\bfr,\bfv)\,
    \txd v_{\perp_1}\,\txd v_{\perp_2}.
\label{sLOSVDs}
\end{equation}
The emissivity is taken into this definition in order to have the
normalization
\begin{equation}
    \int\phi(\bfr,\bfk,v_\parallel)\,\txd v_\parallel
    =
    \frac{\mathcal{L}}{\ell(\bfr)}
    \iiint F(\bfr,\bfv)\,\txd\bfv
    =
    1.
\label{slosvdnorm}
\end{equation}
The spatial LOSVD hence represents the probability for a star at a
position $\bfr$, to have a velocity component $v_\parallel$ in the
direction $\bfk$. The quantity
$\ell(\bfr)\,\phi(\bfr,\bfk,v_\parallel)$ then represents the
stellar emissivity at $\bfr$ of stars which have a velocity
$v_\parallel$ in the direction $\bfk$.

For the calculation of the spatial LOSVD, it is important to know
that a distribution function will not depend on the six phase
space coordinates $(\bfr,\bfv)$ independently, but only through a
number of integrals of motion. In a spherical galaxy, we can write
the distribution function generally as $F(E,L)$, where $E$ and $L$
are the binding energy and angular momentum integrals respectively
(Binney \& Tremaine 1987). Because the distribution function only
depends on $v_\theta$ and $v_\phi$ through the combination
$v_T^2=v_\theta^2+v_\phi^2$, we can choose our reference system
such that $\bfe_{\perp_2}=\bfe_\theta$. If we denote the angle
between $\bfk$ and $\bfe_r$ as $\eta$, we then find
\begin{gather}
    E = \psi(r) - \tfrac{1}{2}v_{\perp_1}^2 -
    \tfrac{1}{2}v_{\perp_2}^2 - \tfrac{1}{2}v_\parallel^2
\label{kin_Exyz}
    \\
    L^2 = r^2\,\left(v_{\perp_1}^2\cos^2\eta
    + v_{\perp_2}^2 + v_\parallel^2\sin^2\eta
    -v_{\perp_1}\,v_\parallel\sin2\eta\right).
\label{kin_Lxyz}
\end{gather}
Substitution of these expressions into the distribution function
allows in principle to calculate the spatial LOSVD using
expression (\ref{sLOSVDs}).

\subsection{The Plummer model}

A major advantage of the Plummer model is it allows the
construction of a completely analytical one-parameter family of
self-consistent dynamical models (Dejonghe~1987). The different
models are characterized by a parameter $q$, restricted by
$q\leq2$. This parameter is determines the orbital structure of
the galaxy: negative values of $q$ correspond to a tangential
anisotropy, positive values to a radial anisotropy, whereas for
$q=0$ the model is isotropic. By means of Laplace-Mellin
transforms the corresponding distribution function $F(E,L)$ can be
calculated analytically for all values of $q$ (Dejonghe 1986). For
general values of $q$, the distribution function can be expressed
in terms of hypergeometric functions. For even integer values of
$q$ however, the distribution function becomes much simpler, which
will allow us to calculate the spatial LOSVD explicitly.

\subsubsection{The isotropic model $q=0$}

For the isotropic dynamical models, the distribution funciton is
only function of the binding energy. For the Plummer model, it is
a simple power law of $E$,
\begin{equation}
    F(E)
    =
    \frac{3}{7\pi^3}\,
    \frac{1}{c^3u^3}\,
    \left(\frac{2E}{u^2}\right)^{7/2},
\label{plum_dfiso}
\end{equation}
where $u=\sqrt{GM/c}$ represents a characteristic velocity.
Substituting the distribution function (\ref{plum_dfiso}) into the
definition (\ref{sLOSVDs}), we obtain, using the expression
(\ref{kin_Exyz}),
\begin{equation}
    \phi(\bfr,\bfk,v_\parallel)
    =
    \frac{256}{63\pi}\,\frac{1}{\sqrt{2\psi(r)}}\,
    \left[1-\frac{v_\parallel^2}{2\psi(r)}\right]^{9/2}.
\label{kin_slosvdiso}
\end{equation}
Because the distribution function of isotropic galaxies is
completely symmetric in the three velocity components, the spatial
LOSVD is independent of the direction $\bfk$. It is
straightforward to check that the normalization condition
(\ref{slosvdnorm}) is satisfied.

\subsubsection{The radial model $q=2$}

For $q=2$, the most radial model in the Plummer family, the
distribution function vanishes for $2E\leq L^2/c^2$, whereas
reduces to a simple form for $2E\geq L^2/c^2$,
\begin{equation}
    F(E,L)
    =
    \frac{3}{4\pi^3}\,\frac{1}{c^3u^5}
    \left(2E-\frac{L^2}{c^2}\right)^{3/2}.
\end{equation}
To calculate the spatial LOSVD, one has to substitute the
expressions (\ref{kin_Exyz}) and (\ref{kin_Lxyz}) into this
distribution function, and integrate it with respect to
$v_{\perp_1}$ and $v_{\perp_2}$, where the integration surface in
determined by the condition $2E\geq L^2/c^2$. After some algebra,
one finds that the LOSVD is given by
\begin{equation}
    \phi(\bfr,\bfk,v_\parallel)
    =
    \frac{16}{5\pi}\,\frac{1}{\sqrt{2A\psi(r)}}\,
    \left[1-\frac{v_\parallel^2}{2A\psi(r)}\right]^{5/2},
\end{equation}
for $|v_\parallel|\geq\sqrt{2A\psi(r)}$, with
\begin{equation}
    A = \frac{r^2\cos^2\eta+c^2}{r^2+c^2} \leq 1.
\end{equation}
The normalization condition (\ref{slosvdnorm}) can be easily
checked.

\subsubsection{Tangential models with $q=-2m$}

For general negative values of $q$, the Plummer distribution
function can be written in terms of hypergeometric series. If $q$
is a negative even integer, $q=-2m$, this series breaks down after
a finite number of terms, and the distribution function then
becomes a finite series, where each term is a power law of $E$ and
$L$ (Dejonghe~1986). The spatial LOSVD can be calculated for each
of these distribution functions following the same recipe as for
the radial $q=2$ model. An alternative, elegant way to calculate
the spatial LOSVDs for these models using Laplace transforms is
presented by De Rijcke~(2000). After lengthy calculations, one
finds eventually
\begin{multline}
    \phi(\bfr,\bfk,v_\parallel)
    =
    \frac{1}{\sqrt{\pi}}\,
    \frac{\Gamma(6+2m)}{\Gamma\left(\frac{11}{2}+2m\right)}\,
    \frac{1}{\sqrt{2\psi(r)}}\,
    \left[1-\frac{v_\parallel^2}{2\psi(r)}\right]^{\frac{9}{2}+2m}
    \\
    \times
    \sum_{j=0}^m
    \frac{\left(\tfrac{1}{2}\right)_j
    \left(-m\right)_j}{(j!)^2}
    \left(\frac{r^2\sin^2\eta}{r^2+c^2}\right)^{2j}
    \left[1-\frac{v_\parallel^2}{2\psi(r)}\right]^{-j}\,
    \\
    \times
    {}_2F_1\left(-j,5+2m;\tfrac{1}{2};
    \frac{v_\parallel^2}{2\psi(r)}\right),
\label{slosvd_tan}
\end{multline}
where $(a)_j$ represents the Pochammer symbol
\begin{equation}
    (a)_j = a(a+1)(a+2)\cdots(a+j-1).
\end{equation}
Setting $q=m=0$ in the expression (\ref{slosvd_tan}), it reduces
to the expression (\ref{kin_slosvdiso}) we found for the isotropic
Plummer galaxy. To check the normalization of the spatial LOSVDs
we have to evaluate the integral
\begin{equation}
    {\mathcal{I}}(m,j)
    =
    \int_0^1 (1-x^2)^{\frac{9}{2}+2m-j}{}_2F_1(-j,2m+5;\tfrac{1}{2};x^2)\,\txd x,
\end{equation}
for all $j\leq m$. By means of formula~(7.512.4) of Gradshteyn \&
Ryzhik~(1965) we find that this integral vanishes for $j>0$, whereas
for $j=0$ we obtain
\begin{equation}
    {\mathcal{I}}(m,0)
    =
    \frac{\sqrt{\pi}}{2}\,
    \dfrac{\Gamma\left(\tfrac{11}{2}+2m\right)}{\Gamma(6+2m)}.
\end{equation}
Only the first term in the summation in equation
(\ref{slosvd_tan}) will hence contribute to the integral over
$v_\parallel$, and it is obvious that the final result will equal
1.

\section{Anisotropic scattering and the importance of the dust grain
velocity terms in equation (\ref{VVV})}

\begin{figure}
\centering
\includegraphics[clip]{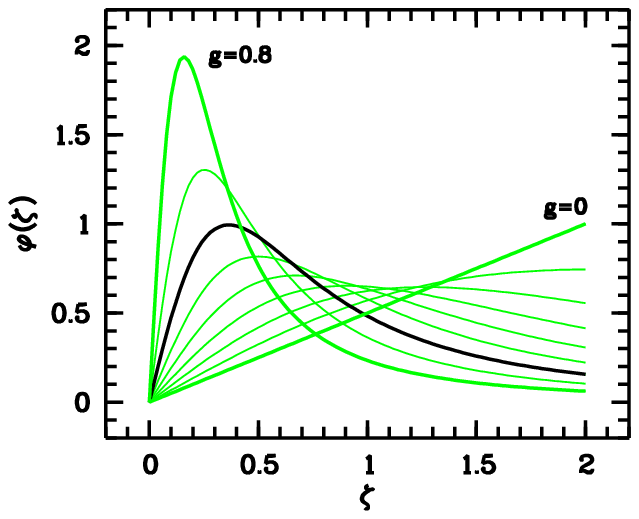}
\caption{The probability density $\varphi(\zeta)$ for
$\zeta\equiv\|\bfk_i-\bfk_{i-1}\|$, corresponding to the
Henyey-Greenstein phase function. Curves are shown for different
values of the asymmetry parameter $g$, ranging between $g=0$ and
$g=0.8$, in steps of 0.1. The black lines indicates the
probability density for $g=0.6$, appropriate for the optical
regime.}
\label{zeta.eps}
\end{figure}

\begin{table}
\centering \caption{The mean, the median and the maximum value for
the probability density $\varphi(\zeta)$ corresponding to the
Henyey-Greenstein phase function, for different values of the
anisotropy parameter $g$.} \label{zeta.tab}
\begin{tabular}{cccc} \hline
$g$ & $\zeta_{\text{mean}}$ & $\zeta_{\text{med}}$ & $\zeta_{\text{peak}}$ \\
\hline
0.0 & 1.33 & 1.41 & 2.00 \\
0.1 & 1.25 & 1.30 & 2.00 \\
0.2 & 1.16 & 1.19 & 1.26 \\
0.3 & 1.07 & 1.06 & 0.90 \\
0.4 & 0.98 & 0.93 & 0.67 \\
0.5 & 0.87 & 0.79 & 0.50 \\
0.6 & 0.75 & 0.64 & 0.37 \\
0.7 & 0.63 & 0.49 & 0.25 \\
0.8 & 0.48 & 0.33 & 0.16 \\
0.9 & 0.29 & 0.17 & 0.07 \\
1.0 & 0.00 & 0.00 & 0.00 \\ \hline
\end{tabular}
\end{table}

Each term in the summation in expression (\ref{VVV}) is a inner
product of the velocity vector $\bfv_{d_i}$ and the relative
direction vector $\bfk_i-\bfk_{i-1}$. It is obvious that the
contribution of one such term does not depend only on the dust
grain velocity $v_{d_i}$, but also on the scattering angle. The
distribution of scattering angles will therefore also play a role
in the importance of the dust grain velocity terms in equation
(\ref{VVV}). If, for example, the propagation direction of a
photon is on average only slightly changed during a scattering
event, we find that $\bfk_i\approx\bfk_{i-1}$, which reduces the
magnitude of the dust grains velocity terms.

We can illustrate this nicely by calculating the statistical
distribution of the modulus of the vector $\bfk_i-\bfk_{i-1}$,
i.e.\ we will calculate the probability density $\varphi(\zeta)$
for $\zeta \equiv \|\bfk_i-\bfk_{i-1}\|$. Since the directions
$\bfk_{i-1}$ and $\bfk_i$ represent unit vectors, elementary
trigonometry learns that
\begin{equation}
    \zeta
    =
    \sqrt{2(1-\cos\alpha)},
\end{equation}
where $\alpha$ is the scattering angle. The scattering angle is
distributed according to the phase function $\Phi(\cos\alpha)$,
such that
\begin{equation}
    \varphi(\zeta)
    =
    \tfrac{1}{2}\,\Phi\left(1-\tfrac{\zeta^2}{2}\right)\,\zeta,
\end{equation}
with $\zeta$ taking values between 0 and 2. If the scattering is
isotropic, the phase function has the simple form
$\Phi(\cos\alpha)=1$, such that $\varphi(\zeta)=\zeta/2$. Larger
values for $\zeta$ will hence be preferred. If we adopt
Henyey-Greenstein scattering (see Section \ref{phafu.sec}), we can
calculate the corresponding values for different degrees of
anisotropy. In Fig.\ \ref{zeta.eps}, the probability density
$\varphi(\zeta)$ corresponding to the Henyey-Greenstein phase
function (\ref{hg}) is shown for different values of anisotropy
parameters $g$. In the limit $g=0$, this function of course
reduces to the isotropic equivalent $\varphi(\zeta)=\zeta/2$. If
the anisotropy parameter increases, however, smaller values for
$\zeta$ will gradually become dominant. We can quantify this trend
by calculating the mean value $\zeta_{\text{mean}}$, the median
$\zeta_{\text{med}}$ or the value $\zeta_{\text{peak}}$ for which
$\varphi(\zeta)$ reaches its maximum value. For the isotropic
phase function we find $\zeta_{\text{mean}}=\tfrac{4}{3}$,
$\zeta_{\text{med}}=\sqrt{2}$ and $\zeta_{\text{peak}} = 2$,
whereas for the Henyey-Greenstein phase function, we obtain
\begin{gather}
    \zeta_{\text{mean}}
    =
    \frac{1-g}{g}
    \left[
    \frac{1+g}{\sqrt{g}}\,
    \ln\left(\frac{1+\sqrt{g}}{\sqrt{1-g}}\right)-1\right]
    \\
    \zeta_{\text{med}}
    =
    \sqrt{g^3-3g+2}
    \\
    \zeta_{\text{peak}}
    =
    \min\left(2,\frac{1-g}{\sqrt{2g}}\right).
\end{gather}
In Table \ref{zeta.tab} we tabulate $\zeta_{\text{mean}}$,
$\zeta_{\text{med}}$ and $\zeta_{\text{peak}}$ as a function of
the asymmetry parameter. This table confirms that smaller values
of $\zeta$ quickly start to dominate as $g$ grows larger. In
particular, for the anisotropy parameters appropriate in the
optical regime, $g\sim0.6$ (see Table \ref{dustprop.tab}), the
density $\varphi(\zeta)$ clearly favours fairly small values for
$\zeta$ (black curve in Fig.\ \ref{zeta.eps}). As a consequence,
$\|\bfk_i-\bfk_{i-1}\|$ will be fairly small in the majority of
the scattering events. The anisotropic nature of scattering hence
contributes to reducing the importance of the dust grain velocity
terms in equation (\ref{VVV}).

\bsp


\begin{thebibliography}{}

\bibitem{AS65} Abramowitz M., Stegun I. A., 1972, Handbook of Mathematical Functions. Dover Publications Inc., New York
\bibitem{BD00} Baes M., Dejonghe H., 2000, MNRAS, 313, 153 [paper~I]
\bibitem{BD01a} Baes M., Dejonghe H., 2001a, MNRAS, 326, 722
\bibitem{BD01b} Baes M., Dejonghe H., 2001b, MNRAS, 326, 733
\bibitem{BD01c} Baes M., Dejonghe H., 2001c, ApJ, 563, L19
\bibitem{BDD00} Baes M., Dejonghe H., De Rijcke S., 2000, MNRAS, 318, 798 [paper~II]
\bibitem{BT89} Bally J., Thronson H. A., 1989, AJ, 97, 69
\bibitem{BPPS93} Bertola F., Pizzella A., Persic M., Salucci P., 1993, ApJ, 416, L45
\bibitem{BFG96} Bianchi S., Ferrara A., Giovanardi C., 1996, ApJ, 465, 127
\bibitem{BDA00} Bianchi S., Davies J. I., Alton P. B., 2000, A\&A, 359, 65
\bibitem{BFDA00} Bianchi S., Ferrara A., Davies J. I., Alton P. B., 2000, MNRAS, 311, 601
\bibitem{BT87} Binney J., Tremaine S., 1987, Galactic Dynamics. Princeton Univ. Press, Princeton
\bibitem{BM82} Blandford R. D., McKee C. F., 1982, ApJ, 255, 419
\bibitem{BBFA92} Bosma A., Byun Y., Freeman K. C., Athanassoula E., 1992, ApJ, 400, L21
\bibitem{CS85} Carlberg R. G., Sellwood J. A., 1985, ApJ, 292, 79
\bibitem{CE59} Cashwell E. D., Everett C. J., 1959, A Practical Manual on the Monte Carlo Method for Random Walk Problems. Pergamon, New York
\bibitem{C34} Chandrasekhar S., 1934, MNRAS, 94, 444
\bibitem{C60} Chandrasekhar S., 1960, Radiative Transfer, Dover Publications, New York
\bibitem{CW95} Code A. D., Whitney B. A., 1995, ApJ, 441, 400
\bibitem{D86} Dejonghe H., 1986, Phys. Rep., 133, 217
\bibitem{D87} Dejonghe H., 1987, MNRAS, 224, 13
\bibitem{D89} Dejonghe H., 1989, ApJ, 343, 113
\bibitem{D00} De Rijcke S., 2000, PhD Thesis, Univ. Gent
\bibitem{DBP95} Di Bartolomeo A., Barbaro G., Perinotto M., 1995, MNRAS, 277, 1279
\bibitem{DDP89} Disney M. J., Davies J. I., Phillips S., 1989, MNRAS, 239, 939
\bibitem{EB85} Ebneter K., Balick B., 1985, AJ, 90, 183
\bibitem{FG79} Faber S. M., Gallagher J. S., 1979, ARA\&A, 17, 135
\bibitem{F70} Freeman K. C., 1970, ApJ, 160, 811
\bibitem{FM00} Ferrarese L., Merritt D., 2000, ApJ, 539, L9
\bibitem{FPPMWJ01} Ferrarese L., Pogge R. W., Peterson B. M., Merritt D., Wandel A., Joseph C. L., 2001, ApJ, 555, L79
\bibitem{FPMC99} Ferrari F., Pastoriza M. G., Macchetto F., Caon N., 1999, A\&AS, 136, 269
\bibitem{FH93} Fich M., Hodge P., 1993, ApJ, 415, 75
\bibitem{FHY94} Fischer O., Henning Th., Yorke H. W., 1994, A\&A, 284, 187
\bibitem{FRR80} Flannery B. P., Roberge W., Rybicki G. B., 1980, ApJ, 236, 598
\bibitem{FJT85} Forman W., Jones C., Tucker W., 1985, ApJ, 293, 102
\bibitem{FvGdZ94} Franx M., van Gorkom J., de Zeeuw T., 1994, ApJ, 293, 102
\bibitem{GBBDFFGGHKLMPRT00a} Gebhardt K., Bender R., Bower G., Dressler A., Faber S. M., Filippenko A. V., Green R., Grillmair C., Ho L. C., Kormendy J., Lauer T. R., Magorrian J., Pinkney J., Richstone D., Tremaine S., 2000a, ApJ, 539, L13
\bibitem{GKHBBDFFGGLMPRT00b} Gebhardt K., Kormendy J., Ho L. C., Bender R., Bower G., Dressler A., Faber S. M., Filippenko A. V., Green R., Grillmair C., Lauer T. R., Magorrian J., Pinkney J., Richstone D., Tremaine S., 2000b, ApJ, 543, 5
\bibitem{G93} Gerhard O. E., 1993, MNRAS, 265, 213
\bibitem{GJSB98} Gerhard O., Jeske G., Saglia R. P., Bender R., 1998, MNRAS, 295, 197
\bibitem{GKSB01} Gerhard O., Kronawitter A., Saglia R. P., Bender R., 2001, AJ, 121, 1936
\bibitem{GKM97} Gerssen J., Kuijken K., Merrifield M. R., 1997, MNRAS, 288, 618
\bibitem{GKM00} Gerssen J., Kuijken K., Merrifield M. R., 2000, MNRAS, 317, 545
\bibitem{GR65} Gradshteyn I. S., Ryzhik I. M., 1965, Table of Integrals, Series and Products. Academic Press Inc., New York
\bibitem{GCIR96} Griffiths R. E., Casertano S., Im M., Ratnatunga K. U., 1996, MNRAS, 282, 1159
\bibitem{GCW98} Gordon K. D., Calzetti D., Witt A. N., 1997, ApJ, 487, 625
\bibitem{GMWC01} Gordon K. D., Misselt K. A., Witt A. N., Clayton G. C., 2001, ApJ, 551, 269
\bibitem{GdJ95} Goudfrooij P., de Jong T., 1995, A\&A, 298, 784
\bibitem{GCS97} Gros M., Crivellari L., Simonneau E., 1997, ApJ, 489, 311
\bibitem{HELTC81} Hawarden T. G., Elson R. A. W., Longmore A. J., Tritton S. B., Corwin H. G. Jr., 1981, MNRAS, 196, 747
\bibitem{HR71} Hummer D. G., Rybicki G. B., 1971, MNRAS, 152, 1
\bibitem{JB90} Jenkins A., Binney J., 1990, MNRAS, 245, 305
\bibitem{J86} Jura M., 1986, ApJ, 306, 483
\bibitem{K00} Kaspi S., Smith P. S., Netzer H., Maoz D., Jannuzi B. T., Giveon U., 2000, ApJ, 533, 631
\bibitem{KGW92} Knapp G. R., Gunn J. E., Wynn-Williams C. G., 1992, ApJ, 399, 76
\bibitem{KKF98} Keeton C. R., Kochanek C. S., Falco E. E., 1998, ApJ, 509, 561
\bibitem{K30} Kosirev N. A., 1934, MNRAS, 94, 430
\bibitem{KSGB00} Kronawitter A., Saglia R. P., Gerhard O., Bender R., 2000, A\&AS, 144, 53
\bibitem{LSR00} Leeuw L. L., Sansom A. E., Robson E., 2000, MNRAS, 311, 683
\bibitem{LW99} Loewenstein M., White R. E. III, 1999, ApJ, 518, 50
\bibitem{MC92} Mathis J. S., Cardelli J. A., 1992, ApJ, 398, 610
\bibitem{MMIRYO98} Matsushita K., Makishima K., Ikebe Y., Rokutanda E., Yamasaki N., Ohashi T., 1998, ApJ, 499, L13
\bibitem{M70} Mattila K., 1970, A\&A, 9, 53
\bibitem{MW01} Matthews L., Wood K., 2001, ApJ, 548, 150
\bibitem{M98} Merluzzi P., 1998, A\&A, 338, 807
\bibitem{MGK01} Merrifield M., Gerssen J., Kuijken K., 2001, in Galaxy Disks and Disk Galaxies, J. G. Funes S. J. and E. M. Corsini eds., ASP Conf. Series, 230, 221
\bibitem{M00} Michard R., 2000, A\&A, 360, 85
\bibitem{M78} Mihalas D., 1978, Stellar Atmospheres, W. H. Freeman and Co., San Francisco
\bibitem{MGCW01} Misselt K. A., Gordon K. D., Clayton G. C., Wolff M. J., 2001, ApJ, 551, 277
\bibitem{PV85} Peraiah A., Varghese B. A., 1985, ApJ, 290, 411
\bibitem{P11} Plummer H. C., 1911, MNRAS, 71, 460
\bibitem{PBMCG94} Prada F., Beckman J. E., McKeith C. D., Castles J., Greve A., 1994, ApJ, 423, L35
\bibitem{PFTV89} Press W. H., Flannery B. P., Teukolsky S. A., Vetterling W. T., 1989, Numerical Recipes. Cambridge Univ. Press, Cambridge
\bibitem{RdZCvdMC97} Rix H.-W. de Zeeuw P. T., Cretton N., van der Marel R. P., Carollo C. M., 1997, ApJ, 488, 702
\bibitem{RM84} Rogers C., Martin P. G., 1984, ApJ, 284, 327
\bibitem{R80} Rowan-Robinson M., 1980, ApJS, 44, 403
\bibitem{SBS92} Saglia R. P., Bertin G., Stiavelli M., 1992, ApJ, 384, 433
\bibitem{SBBDDSSdZZ93} Saglia R. P., Bertin G., Bertola F., Danziger J., Dejonghe H., Sadler E. M., Stiavelli M., de Zeeuw P. T., Zeilinger W. W., 1993, ApJ, 403, 567
\bibitem{S75} Schmid-Burgk J., 1975, A\&A, 40, 249
\bibitem{SS51} Spitzer L. Jr., Schwarzschild M., 1951, ApJ, 114, 385
\bibitem{SS53} Spitzer L. Jr., Schwarzschild M., 1953, ApJ, 118, 106
\bibitem{TAWTI00} Tomita A., Aoki K., Watanabe M., Takata T., Ichikawa S., 2000, AJ, 120, 123
\bibitem{TTFDJvR01} Tran H. D., Tsvetanov Z., Ford H. C., Davies J., Jaffe W., van den Bosch F. C., Rest A., 2001, AJ, 121, 2928
\bibitem{vdMF93} van der Marel R. P., Franx M., 1993, ApJ, 407, 525
\bibitem{vDF95} van Dokkum P. G., Franx M., 1995, AJ, 110, 2027
\bibitem{VV88} V\'{e}ron-Cetty M.-P., V\'{e}ron P., 1988, A\&A, 204, 28
\bibitem{Wi77} Wielen R., 1977, A\&A, 60, 263
\bibitem{WH95} Wiklind T., Henkel C., 1995, A\&A, 297, L71
\bibitem{WS96} Wise M. S., Silva D. R., 1996, ApJ, 461, 155
\bibitem{W77} Witt A. N., 1977, ApJS, 35, 1
\bibitem{WG96} Witt A. N., Gordon K. D., 1996, ApJ, 463, 681
\bibitem{WBS84} Witt A. N., Bohlin R. C., Stecher T. P., 1984, ApJ, 279, 698
\bibitem{WTC92} Witt A. N., Thronson H. A. Jr., Capuano J. M. Jr., 1992, ApJ, 393, 611
\bibitem{WFP98} Wolf S., Fischer O., Pfau W., 1998, A\&A, 340, 103
\bibitem{WHS99} Wolf S., Henning Th., Stecklum B., 1999, A\&A, 349, 839
\bibitem{WR99} Wood K., Reynolds R. J., 1999, ApJ, 525, 799
\bibitem{WD96} Wybo M., Dejonghe H., 1996, A\&A, 312, 649
\bibitem{Y80} Yorke H. W., 1980, A\&A, 86, 286
\bibitem{YMW84} Yusef-Zadeh F., Morris M., White R. L., 1984, ApJ, 278, 186


\end{thebibliography}
\end{document}